\documentclass[aps, prl, a4paper, reprint, showkeys, floatfix]{revtex4-1}
\usepackage{dcolumn}
\usepackage{lipsum}
\usepackage{hyperref} 
\hypersetup{
     colorlinks   = true,
     linkcolor	=	black,
     urlcolor	=	black,
     citecolor    = black
}
\usepackage{fancyhdr}
\usepackage{mathtools}
\usepackage{multirow}
\usepackage{verbatim}
\usepackage{amsthm}
\usepackage{amsmath}
\usepackage{amssymb}
\usepackage{graphicx}
\usepackage{float}
\usepackage[T1]{fontenc}
\usepackage{fancyref}
\usepackage{etoolbox}
\usepackage[caption=false]{subfig}
\usepackage{color}
\usepackage{tikz}
\usetikzlibrary{backgrounds}
\usetikzlibrary{graphs}
\usetikzlibrary{graphs.standard}
\usetikzlibrary{shapes.misc}
\begin{document}

\title{A local graph rewiring algorithm for sampling spanning trees}
\author{Neal~McBride}
\email{neal.mcbride@tcd.ie}
\affiliation{CONNECT / The Centre for Future Networks and Communications, Trinity College Dublin, Ireland}
\author{John~Bulava}
\affiliation{CP$^{3}$-Origins, University of Southern Denmark, Campusvej 55, 5230 Odense M, Denmark}

\date{\today}


\begin{abstract}
We introduce a Markov Chain Monte Carlo algorithm which samples from the space of spanning trees of complete graphs using local rewiring operations only. The probability distribution of graphs of this kind is shown to depend on the symmetries of these graphs, which are reflected in the equilibrium distribution of the Markov chain. We prove that the algorithm is ergodic and proceed to estimate the probability distribution for small graph ensembles with exactly known probabilities. The autocorrelation time of the graph diameter demonstrates that the algorithm generates independent configurations efficiently as the system size increases. Finally, the mean graph diameter is estimated for spanning trees of sizes ranging over three orders of magnitude. The mean graph diameter results agree with theoretical asymptotic values.

\vspace{4mm}
\noindent
\textit{Preprint number: CP3-Origins-2017-032 DNRF90}

\end{abstract}
\maketitle

\section{Introduction} \label{sec:intro}

Physical processes which can be described using graph theory often involve networks which evolve over time. For example, in wireless telecommunications, the edges between nodes which interfere can change due to network load or failure. In biological systems, the network of interactions between genes, proteins and other biologically relevant molecules change due to the current needs of the cell \cite{Luscombe:04}. 

Studying the properties of evolving networks usually involves taking a `snapshot' of the physical network that is thought to be `characteristic' of its general behavior. Some properties of the network are then studied and in some cases this is repeated for a number of snapshots. However, this approach forbids us from asking questions about `equilibrium' properties of a network ensemble, given some constraints such as the number of nodes ($\lvert V\rvert$) or edges ($\lvert E\rvert$) in a graph. 

In this paper, we describe an approach to study network ensembles through rewiring using a new Markov chain Monte Carlo method. We propose a new algorithm to explore the configuration space of a network ensemble. Our rewiring algorithm performs `local' rewiring updates by adding and removing edges close to each other. This procedure conserves the number of nodes and edges in the network. In the following sections, we provide the motivation behind our algorithm. The algorithm is then presented and proven to be ergodic. Our estimate of the probability distribution of small graphs as defined in Eqs.~\ref{eq:labelling} and \ref{eq:eq_dist} and an observable called the graph diameter is shown to be consistent with exact values. The graph diameter is shown to scale like the square root of the number of nodes and the autocorrelation time of our algorithm also scales slowly with system size. This slow increase in autocorrelation time with system size demonstrates that our algorithm can sample efficiently even for large graphs. 

\section{Background \label{sec:background}}
One of the earliest and most well-known network rewiring models is the Watts-Strogatz model introduced in Ref.~\cite{Watts:98}. This random graph construction model performs a rewiring of the edges incident to each node with a given probability which is a parameter of the model. The resulting graph has properties in between a regular lattice and a fully random graph. The target node of each rewired edge is chosen with uniform probability from the other nodes. This random graph model was introduced as the simplest possible method to build graphs which exhibit a high amount of clustering and small average path length compared to a random graph. This is a significant result, since the `small-world phenomenon' described by these features is characteristic of very well-studied real-world networks. These two phenomena are present in networks such as the U.S. power-grid, the neural network of \textit{C. elegans} and the network of movie actors who have worked together \cite{Barabasi:02}. However, the Watts-Strogatz model produces a single static graph and does not attempt to explore the entire graph configuration space.

To the best of our knowledge, the first Monte Carlo algorithm for the purpose of rewiring graph adjacency matrices appears in Ref.~\cite{Ramachandra:96}. The algorithm randomly selects four matrix elements, $(i_1j_1, i_1j_2, i_2j_2, i_2j_1 )$, which form a rectangular cycle in the adjacency matrix. The adjacent cycle entries must take alternating values, for example $(1, 0, 1, 0)$, where the `$1$'s represent the presence of an edge $\{i, j\}$. The values of the adjacent cycle entries are then exchanged, proposing a new Monte Carlo state. Since the entries form a rectangle in the matrix, the sum over the rows and columns is invariant under this transformation and the degree distribution is preserved. 

The number of rewires necessary for degree distribution preserving algorithms to produce independent graphs is not currently known, but a number of estimates have been given. A similar rewiring algorithm to Ref.~\cite{Ramachandra:96} is proposed in Ref.~\cite{Kannan:99} which accepts all proposed transitions. The procedure is known as the degree distribution (DD) preserving algorithm and is performed using the steps outlined below.
\begin{itemize}
\item{Pick a node $u_1$ from all nodes in a graph with uniform probability and one of its neighbors $u_2$. This set of nodes is the first edge.}
\item{Pick a second node, $v_1$ and uniformly choose an incident neighbor called $v_2$.}
\item{Swap the edges ($u_1, u_2$) and ($v_1, v_2$) to create ($u_1, v_2$) and ($v_1, u_2$).}
\end{itemize}
The authors in Ref.~\cite{Kannan:99} show that the minimum relaxation time of the algorithm is at least $O(\lvert E\rvert^6)$. However, the bound is very wide and depends on an arbitrarily chosen precision $\epsilon$: the distance between the true and observed stationary distributions. 

In Ref.~\cite{Stanton:12}, a joint degree distribution (JDD) preserving algorithm is presented which samples graphs with a joint probability matrix, $K(i, j)$, of the number of edges between vertices of degree $i$ and $j$. The following steps perform the JDD preserving sampling.
\begin{itemize}
\item{Pick an endpoint $u_1$ with uniform probability and a uniform random neighbor $u_2$. This is the first edge.}
\item{Pick a second endpoint, $v_1$, such that the degrees $k(u_1) = k(v_1)$ and uniformly choose an incident neighbor called $v_2$.}
\item{Swap the edges ($u_1, u_2$) and ($v_1, v_2$) to create ($u_1, v_2$) and ($v_1, u_2$).}
\end{itemize}
The authors do not propose a theoretical bound on convergence.

Since a tight theoretical bound on the number of rewires necessary to produce independent graphs does not currently exist for either of these algorithms, a recent paper has proposed an effective rule of thumb which can be used instead~\cite{Ray:15}. The number of necessary rewires is estimated using the stationary property of the graph probability distribution and a chosen precision level $\epsilon$, between the true and observed distributions. The authors prove that a graph with approximately independent edges and a specified degree distribution can be generated after $N_{\mathrm{ind}} = \frac{1}{2}\lvert E \rvert \ln \epsilon^{-1}$ rewires. The same approximation for the JDD preserving algorithm requires $N_{\mathrm{ind}} = \lvert E \rvert \ln \epsilon^{-1}$. In a  graph containing roughly $10^3$ nodes, $7.5 \lvert E \rvert$ and $15 \lvert E \rvert$ rewires are considered sufficient for the DD and JDD preserving algorithms respectively. For practical Monte Carlo simulations, this is an acceptable number of rewires for graphs of this size. The majority of edges become independent after this number of rewires, but a small number of edges are particularly persistent and need an order of magnitude more rewires to become independent.

The algorithms mentioned so far have used rewiring for graph formation or as a Monte Carlo method to sample the space of graphs with a given degree distribution or joint degree distribution. To the best of our knowledge, there have been no published examples of a Monte Carlo algorithm which samples from the space of graphs with a constant number of nodes and edges. Graphs of this kind can occur in sensor networks, where a number of sensors form a network to communicate, which can change over time~\cite{Abbasi:07}. Furthermore, it may be necessary to estimate ensemble averages of observables which depend on both the graph structure and some other dynamical element embedded on the graph, for example a Potts spin \cite{McBride:2017}. With this in mind, we present a method to sample from the ensemble of graphs with a given $|E|$ and $|V|$. This work focuses on the most basic graph ensemble; spanning trees of the complete graph $K_n$, which defines $|E| = n-1$ and $|V| = n$. Spanning trees are connected and contain no loops, repeated edges or self-loops. Spanning trees are an important class of graphs, since they contain the minimum number of edges to keep the graph connected and there exists only one path between any pair of nodes.

\section{Motivating the Algorithm} \label{sec:algorithm}
\begin{figure} 
	\centering
	\begin{tikzpicture}
			\tikzstyle{every node} = [draw, shape=circle, minimum size = 0.9cm]
 			\graph { subgraph K_n [n=5,clockwise, radius = 1.5cm] };
	\end{tikzpicture}
	\caption{The $K_{5}$ graph is a complete graph with 5 vertices. Our algorithm samples from the configuration space of spanning trees of complete graphs. These are tree graphs which contain every vertex and the minimum number of edges.}
	\label{graph:K_5}
\end{figure}
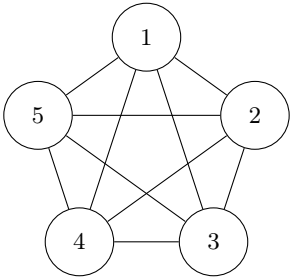

Our Monte Carlo algorithm samples from the configuration space of graphs with a fixed number of nodes and edges. This sampling depends on the graph probability distribution, $\pi_g$. The set of graphs of interest are spanning trees of a complete graph. For example, the complete graph with five nodes, $K_5$, is shown in Fig.~\ref{graph:K_5}. A spanning tree of $K_5$ is a subgraph that contains five nodes and four of the edges in $K_5$. Later, we show that the graph probability distribution depends on the symmetries of the subgraphs.

The graph probability distribution can be used to estimate the ensemble average of observables which depend on graph structure;
\begin{align*}
\langle O \rangle = \sum_{g \in G} O(g) \pi_g.
\end{align*}
We use the notation throughout this chapter that $G$ is the ensemble of graphs and $g$ is a particular graph configuration. 
The graph observable of particular interest in this paper is the diameter ($d$). In tree graphs, the diameter is the longest path. The diameter is very interesting because it depends on the entire structure of the graph, is bounded by the number of edges and is easily calculated. Since it depends on the overall graph structure, the diameter is a slowly changing observable compared to other graph observables. The integrated autocorrelation time ($\tau_{\mathrm{int}}$) of the diameter can therefore be used to judge the efficiency of our Monte Carlo algorithm. This allows us to measure the number of rewires necessary to sample independent graphs. It should be noted however, that performing $\tau_{\mathrm{int}}$ rewiring sweeps does not guarantee that all edges have been rewired. Our estimates of $\tau_{\mathrm{int}}$ are calculated using the method outlined in Ref.~\cite{Wolff:04}.

For practical purposes, it is necessary to consider each node with an identifying label in order to describe it unambiguously. Cayley's formula states that the number of spanning trees of a complete graph containing $n$ nodes, with a label at each vertex, is $n^{n-2}$. Many of these graphs will have the same structure of adjacency and non-adjacency of nodes. Graphs with this property are isomorphic. Graph isomorphism defines an equivalence class of graphs, which may differ by their labelling structure. 

We now give some necessary definitions which can be found in any textbook on graph theory, e.g. Ref.~\cite{Trudeau:2013}. Two simple graphs $G$ and $H$ are isomorphic if there is a bijection,
\begin{align*}
\Theta: V(G) \rightarrow V(H),
\end{align*}
which preserves adjacency and non adjacency of vertices. Therefore each edge must obey,
\begin{align} \label{eq:iso_edges}
\{u, v\} \in E(G) \iff \{\Theta(u), \Theta(v)\} \in E(H),
\end{align}
where $\{u, v\}$ is an edge in the graph $G$ and $\{\Theta(u), \Theta(v)\}$ is the edge mapped to the graph $H$. Both graphs are isomorphic, $G \cong H$,  if Eq.~\ref{eq:iso_edges} is satisfied for all edges in the edge sets $E(G)$ and $E(H)$. 

An automorphism is a permutation of the vertex set which preserves adjacency and non-adjacency between the graph and its image. It maps G to itself. Therefore, an automorphism, $\alpha$, of a graph is an isomorphism between $G$ and itself,
\begin{align*}
\alpha:V(G) \rightarrow V(G),
\end{align*}
such that
\begin{align*}
\{\alpha(u), \alpha(v)\} \in E(G) \iff \{u, v\} \in E(G).
\end{align*}
\begin{figure}
	\centering
	\subfloat[Linear graph.]{
	\begin{tikzpicture}
		\tikzstyle{every node} = [draw, shape=circle, minimum size=0.9cm]
		\node(v1) at (0,0) {};
		\node(v2) at (1.5,0) {};
		\node(v3) at (3,0) {};
		\node(v4) at (4.5,0) {};
		\node(v5) at (6,0) {};
		\draw (v1) -- (v2);
		\draw (v2) -- (v3);
		\draw (v3) -- (v4);
		\draw (v4) -- (v5);
	\end{tikzpicture}
	\label{graph:Line_5}
	}\\ \bigskip
	\subfloat[Fork graph.]{
	\begin{tikzpicture}
		\tikzstyle{every node} = [draw, shape=circle, minimum size=0.9cm]
		\node(v1) at (0,0) {};
		\node(v2) at (1.5,0) {};
		\node(v3) at (3,0) {};
		\node(v4) at (3.75,1.3) {};
		\node(v5) at (3.75,-1.3) {};
		\draw (v1) -- (v2);
		\draw (v2) -- (v3);
		\draw (v3) -- (v4);
		\draw (v3) -- (v5);
	\end{tikzpicture}
	\label{graph:Fork_5}
	}\\ \bigskip
	\subfloat[Star graph.]{
	\centering
	\begin{tikzpicture}
		\tikzstyle{every node} = [draw, shape=circle, minimum size=0.9cm]
		\node(v1) at (0,0) {};
		\node(v3) at (1.5,0) {};
		\node(v4) at (0,-1.5) {};
		\node(v5) at (-1.5,0) {};
		\node(v2) at (0,1.5) {};
		\draw (v1) -- (v2);
		\draw (v1) -- (v3);
		\draw (v1) -- (v4);
		\draw (v1) -- (v5);
	\end{tikzpicture}
	\label{graph:Star_5}
	}
	\caption{The three graph isomorphism classes of spanning trees of $K_5$. We have named the classes in order to refer to them with relative ease. In most cases, these are not commonly accepted graph names and are chosen to be descriptive.}
\end{figure}
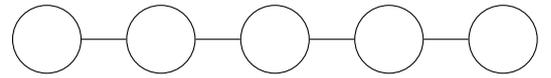
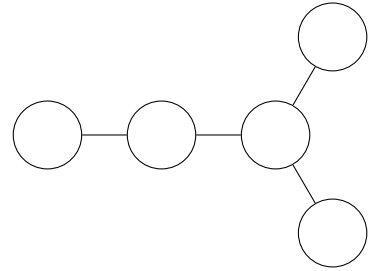
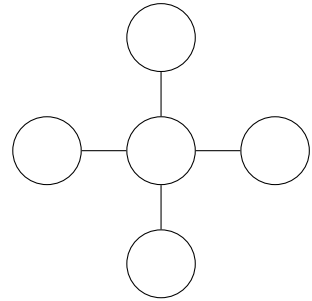
The set of automorphisms of a graph obey the following conditions:
\begin{itemize}
\item{If $\alpha$ and $\beta$ are automorphisms of a graph, then their composition $\alpha \circ \beta$ is also an automorphism.}
\item{Composition of automorphisms is always associative.}
\item{The identity map is always an automorphism of a graph.}
\item{Since automorphisms are bijections, if $\alpha$ is an automorphism of a graph, then the inverse, $\alpha^{-1}$ exists and is also an automorphism.}
\end{itemize}
Since the set of automorphisms obey closure, associativity, identity and inverse, they form a group under composition of morphisms. The automorphism group of a graph is denoted $\mathrm{Aut}(g)$. 

For small complete graphs, it is possible to identify and count all of the isomorphic spanning trees. Many of the isomorphic graphs share the same assignment of labels, or labelling, but differ by a symmetry operation. Such graphs are automorphic to each other.

The probability distribution of spanning trees depends on the number of graph labellings of each isomorphism class. The number of labellings of a class is 
\begin{equation} \label{eq:labelling}
l(g_i) = \frac{n!}{\lvert \mathrm{Aut}(g_i)\rvert},
\end{equation}
where $\lvert \mathrm{Aut}(g_i)\rvert$ describes the size of the automorphism group of the class $g_i$. Summing over the number of labellings of each isomorphic graph class of spanning trees gives the size of the ensemble:
\begin{equation*} \label{eq:sum_relabels}
\sum\limits_{g_i \in G} l(g_{i}) = n^{n-2}.
\end{equation*}
In practice, the equivalence classes $g_i$ may be distinguished for small graphs by some convenient graph invariant e.g. max degree. This can be seen in Figs.~\ref{graph:Line_5}, \ref{graph:Fork_5} and \ref{graph:Star_5}. For larger graphs, this becomes a far more involved task.

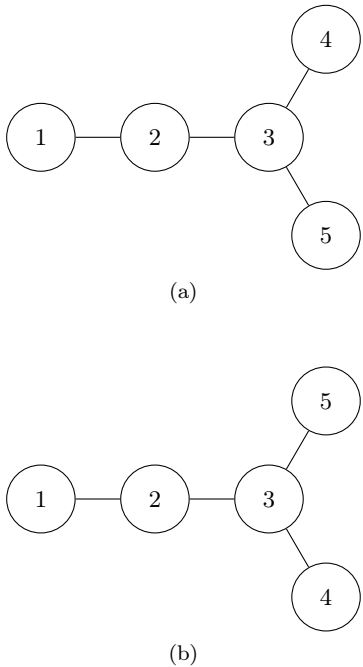
\begin{figure} 
	\centering
	\subfloat[]{
		\begin{tikzpicture} \label{graph:Fork_a}
			\tikzstyle{every node} = [draw, shape=circle, minimum size = 0.9cm]
			\node(v1) at (0,0) {1};
			\node(v2) at (1.5,0) {2};
			\node(v3) at (3,0) {3};
			\node(v4) at (3.75,1.3) {4};
			\node(v5) at (3.75,-1.3) {5};
			\draw (v1) -- (v2);
			\draw (v2) -- (v3);
			\draw (v3) -- (v4);
			\draw (v3) -- (v5);
		\end{tikzpicture}
	} \\ \bigskip
	\subfloat[]{
		\begin{tikzpicture} \label{graph:Fork_b}
			\tikzstyle{every node} = [draw, shape=circle, minimum size = 0.9cm]
			\node(v1) at (0,0) {1};
			\node(v2) at (1.5,0) {2};
			\node(v3) at (3,0) {3};
			\node(v4) at (3.75,1.3) {5};
			\node(v5) at (3.75,-1.3) {4};
			\draw (v1) -- (v2);
			\draw (v2) -- (v3);
			\draw (v3) -- (v4);
			\draw (v3) -- (v5);
		\end{tikzpicture}
	}
	\caption{The two automorphisms of the Fork graph in Fig.~\ref{graph:Fork_5}. Fig.~\ref{graph:Fork_a} is the result of the identity and Fig.~\ref{graph:Fork_b} is the result of a reflection.}
	\label{graph:Fork_automorphisms}
\end{figure}

Conceptually, the automorphism group is the group of symmetries of a graph. A very symmetric graph has a large automorphism group. Complete graphs are the most symmetric by definition. The size of the automorphism group of a complete graph is $n!$, the proof of which is quite simple. Any permutation of the vertices will preserve the adjacency and non-adjacency of the graph, which means that the group has the same size as the set of permutations of $n$ elements. Therefore, there are $n!$ unique permutations of $K_n $ which are all automorphisms:
\begin{align*}
\lvert \mathrm{Aut}(K_n) \rvert = n!.
\end{align*}

The stationary distribution of graphs $\pi_g$ is governed by the rules of probability:
\begin{align*}
\pi_g &\in [0,1]\:, \hspace{3mm} \forall\: g \in G \\
\pi_g &= \left(n^{n-2}\right)^{-1} \\
\sum\limits_{g \in G} \pi_g &= 1
\end{align*}
The probability distribution of the equivalence class $g_i$ is therefore given by
\begin{align*}
\pi_{g_i} = \sum_{g \in g_i} \pi_g.
\end{align*}
Given that we know the number of labeled graphs in each equivalence class from Eq.~\ref{eq:labelling}, the class probability which defines our ensemble is
\begin{align}\label{eq:eq_dist}
\pi_{g_i} = \frac{l(g_i)}{n^{n-2}}.
\end{align}

Before describing the graph rewiring algorithm, we will show how to calculate the exact probability distribution for spanning trees of $K_5$. The three equivalence classes of spanning trees of $K_5$, distinguished by their maximum degree are the Linear, Fork and Star graphs. The size of their respective automorphism groups are
\begin{align*}
\lvert \mathrm{Aut}(g_{\mathrm{Line}})\rvert &= 2,\\
\lvert \mathrm{Aut}(g_{\mathrm{Fork}})\rvert &= 2, \\
\lvert \mathrm{Aut}(g_{\mathrm{Star}})\rvert &= (n-1)! = 24.
\end{align*}
The size of the automorphism groups of the Line and Fork graph can be easily determined by inspecting the morphisms which induce graph symmetries and including the identity morphism. The automorphism group of the star is not as trivial to determine. One way of determining this is to use the relation that the automorphism group of a graph is the same as its graph complement. The complement of a graph is obtained by removing all edges and placing edges where none existed. The complement of the Star graph is the union of $K_4$ and the cycle graph with one node, $C_1$; all radial nodes are adjacent to each other and the central node is a disconnected component of size one. The automorphism group of this graph is isomorphic to that of $K_4$, namely $4!$. Therefore, the star graph with five nodes, $S_4$, has an automorphism group of size $4!$. The number of labellings of each graph equivalence class of spanning trees of $K_5$ can now be given as
\begin{align*}
l(g_{\mathrm{Line}}) &= n!/2 = 60,\\
l(g_{\mathrm{Fork}}) &= n!/2 = 60,\\
l(g_{\mathrm{Star}}) &= n!/24 = 5.
\end{align*}
This demonstrates that the sum of labellings of the three equivalence classes of spanning trees of $K_5$ account for all of the labeled spanning trees;
\begin{align*}
l(g_{\mathrm{Line}}) + l(g_{\mathrm{Fork}}) + l(g_{\mathrm{Star}}) = n^{n-2}.
\end{align*}

\section{Algorithm Definition} \label{sec:rewiring}
In this section, we outline our Monte Carlo rewiring procedure and prove that the resulting Markov chain is ergodic. Therefore, we must establish that the Markov chain produced by the algorithm is irreducible, aperiodic, and positive recurrent. Following this, we discuss the large-graph limit of the graph diameter.

\subsection{Rewiring Steps}
The steps necessary to perform an update sweep of the graph are
\begin{itemize}\label{alg:rewiring}
	\item{Pick an edge, $e$, randomly from the set of all edges with uniform probability. The nodes incident to this edge are the head $(H)$ and tail $(T)$ nodes, which are distinguished later. }
	\item{Form a set of the nodes in the neighborhood of both $H$ and $T$ called $S = \{(N(H) \cup N(T)\} \setminus \{H, T\}$, where $N(V)$ is the set of vertices adjacent to a vertex $V$.}
	\item{Pick one node, $M$, with uniform probability from $S$ as the node to be moved.}
	\item{Determine the head ($H$) and tail ($T$) from the endpoints of $e$ so that $M \in N(H)$.}
	\item{Remove the edge between  $H$ and $M$, ($H \nsim M$).}
	\item{Create an edge between $T$ and $M$, ($T \sim M$).}
\end{itemize} 
The initial configuration of the Markov chain is usually chosen to be the Linear graph such as in Fig.~\ref{graph:Line_5} for convenience. This initial configuration has no detrimental effect to our sampling because our Markov chain is universal and we carefully monitor the necessary thermalisation time before taking any measurements.

The number of edges in the graph remains constant as one edge is created and destroyed per rewire. Since we are sampling the ensemble of spanning trees with $n$ nodes and $n-1$ edges, no cycles can be produced solely from a rewiring operation in this ensemble. The spanning trees of $K_n$ are all connected, since there exists a path between every pair of nodes. Our algorithm preserves this connectivity by splitting the graph into two disconnected subgraphs when $H \nsim M$ and reconnecting them with $T \sim M$.

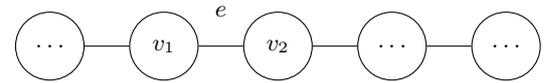
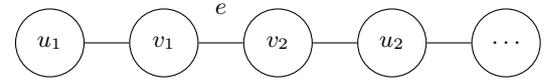
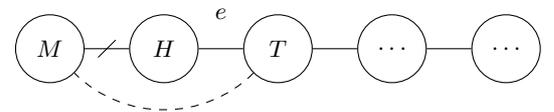
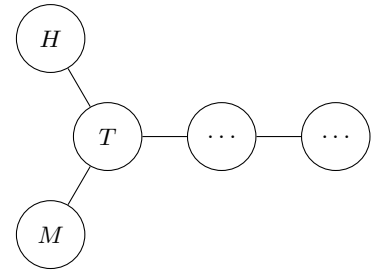
\begin{figure}
	\centering
	\subfloat[Step 1: Choose $e$ with uniform probability $p_e = \lvert E \rvert^{-1}$, from all edges in the graph.]{
		\begin{tikzpicture} \label{graph:rewire_stepa}
			\tikzstyle{every node} = [draw, shape=circle, minimum size=0.9cm];
			\node(v1) at (0,0) {$\ldots$};
			\node(v2) at (1.5,0) {$v_1$};
			\node(v3) at (3,0) {$v_2$};
			\node(v4) at (4.5,0) {$\ldots$};
			\node(v5) at (6,0) {$\ldots$};

			\draw (v1) -- (v2);
			\draw (v2) -- (v3) node[fill=none, draw=none, midway, above] {$e$};
			\draw (v3) -- (v4);
			\draw (v4) -- (v5);
		\end{tikzpicture}
	}\bigskip \\
	\subfloat[Step 2 \& 3: Determine the neighbors of endpoints $v_1$ and $v_2$. In this case, they are labeled $u_1, u_2, v_1, v_2$. The set $S$ is the union of these, excluding $v_1$ and $v_2$ themselves: $S = \{N(v_1) \cup N(v_2)\}\setminus\{v_1, v_2\}$. Pick the node to be moved, $M$, from $S$ with uniform probability $p_M = \lvert S \rvert^{-1}$, where $\lvert S \rvert$ is the number of element in $S$.]{
		\begin{tikzpicture} \label{graph:rewire_stepb}
			\tikzstyle{every node} = [draw, shape=circle, minimum size=0.9cm];
			\node(v1) at (0,0) {$u_1$};
			\node(v2) at (1.5,0) {$v_1$};
			\node(v3) at (3,0) {$v_2$};
			\node(v4) at (4.5,0) {$u_2$};
			\node(v5) at (6,0) {$\ldots$};

			\draw (v1) -- (v2);
			\draw (v2) -- (v3) node[fill=none, draw=none, midway, above] {$e$};
			\draw (v3) -- (v4);
			\draw (v4) -- (v5);
		\end{tikzpicture}
	}\bigskip \\
	\subfloat[Step 4, 5 \& 6: Determine the head node $H$ to be the endpoint of the edge $e$ which is incident to the node $M$, or $H \sim M$. Destroy the edge between $H$ and $M$, $H \nsim M$. Create a new edge between $T$ and $M$,  $T \sim M$.]{
		\begin{tikzpicture} \label{graph:rewire_stepc}
			\tikzstyle{every node} = [draw, shape=circle, minimum size=0.9cm];
			\node(v1) at (0,0) {$M$};
			\node(v2) at (1.5,0) {$H$};
			\node(v3) at (3,0) {$T$};
			\node(v4) at (4.5,0) {$\ldots$};
			\node(v5) at (6,0) {$\ldots$};

			\draw (v1) -- node[strike out, draw, -, minimum size=0.2cm]{}(v2);
			\draw (v2) -- (v3) node[fill=none, draw=none, midway, above] {$e$};
			\draw (v3) -- (v4);
			\draw (v4) -- (v5);
			\draw[dashed] (v3) to[out=225,in=-45] (v1);
		\end{tikzpicture}
	}\bigskip \\
	\subfloat[This is the rewired graph after one iteration of our algorithm. Having performed one rewire, a graph-dependent observable is calculated and stored as the Markov chain history and for analysis at a later stage.]{
		\begin{tikzpicture} \label{graph:rewire_stepd}
			\tikzstyle{every node} = [draw, shape=circle, minimum size = 0.9cm]
			\node[draw=none](v0) at (-1.125,0) {};
			\node(v1) at (0,1.3) {$H$};
			\node(v2) at (0,-1.3) {$M$};
			\node(v3) at (0.75,0) {$T$};
			\node(v4) at (2.25,0) {$\ldots$};
			\node(v5) at (3.75,0) {$\ldots$};
			\node[draw=none](v6) at (5,0) {};

			\draw (v1) -- (v3);
			\draw (v2) -- (v3);
			\draw (v3) -- (v4);
			\draw (v4) -- (v5);
		\end{tikzpicture}
	}
	
	\caption{An iteration of our graph rewiring Markov chain Monte Carlo algorithm on spanning trees of $K_5$.}\label{graph:rewire_step_5}
\end{figure}
When rewiring the node $M$, the entire subgraph attached to it is rewired to $T$. Choosing a node with degree $k=1$ for $M$ results in only a single node being translated along the graph. All rewires are essentially a translation of the subgraph containing $M$ but not $H$. If the chosen edge $e$ has a degree $k = 1$ node as an endpoint, this node will be incorporated into the graph. The degree of the head node always decreases and the degree of the tail node always increases upon successful completion of a rewire. In order to lower the degree of a node, it must be the head of a rewire step. The $H$ vertex must have more than one neighbor (one neighbor other than $T$), otherwise the set $S$ contains only neighbors of $T$. Fig~\ref{graph:rewire_step_5} illustrates how the degree of the $T$ node always increases while that of $H$ always decreases. 

Having defined the steps involved in the rewiring algorithm, it is necessary to prove that the resulting Markov chain converges to the desired probability distribution. The necessary properties to ensure convergence are discussed in the next section. It is also proven that the algorithm possesses these properties. These proofs are non-trivial, as our algorithm does not generally have a finite probability to `do nothing' like the Metropolis algorithm \cite{Metropolis:53}.

\subsection{Ergodicity}

Our rewiring algorithm produces a series of random graphs in which each successive graph is dependent on the previous graph. The probability of realizing a graph $y$ at time $t+1$, given graph $x$ at time $t$ is the transition probability:
\begin{equation*}
P(X_{t+1} = y| X_{t} =x) = p_{xy}.
\end{equation*}  
By construction, our algorithm satisfies this Markov property and the resulting graph sequence is a Markov chain.

A Markov chain is ergodic if it is irreducible, aperiodic and positive recurrent~\cite{Sokal1996}. The fundamental limit theorem for irreducible Markov chains states that for any ergodic Markov chain, there exists a unique stationary distribution:
\begin{align*}
\pi_y = \lim_{n \to \infty} p^{(n)}_{xy},
\end{align*}
where $p^{(n)}_{xy}$ is the $n$-step transition probability
\begin{align*}
p^{(n)}_{xy} = P(X_{t+n} = y| X_{t} =x).
\end{align*}

In order to estimate $\pi$-averaged observables by time-averaging over the successive graph configurations produced by our algorithm, we must prove that it is ergodic. We start by considering the accessibility of every pair of Markov chain states.

\subsubsection{Irreducibility}
A Markov chain is irreducible if for each pair of states, ($x, y$), there exists an $n \geq 0$ for which $p^{(n)}_{xy} > 0$. We must demonstrate that it is possible to transition from each graph state to every other graph state for our algorithm to produce an irreducible Markov chain. It is sufficient to show that every state can access the Linear graph and return, even if this transition occurs through many intermediate states.
 
Each vertex in a Linear graph has degree $k(v_i) =2$, except the endpoints which have degree $k(v_{\mathrm{end}})=1$. All other states in the space $G$ of spanning trees of the complete graph $K_n$ for $n > 3$ must have at least one vertex with a degree larger than two. Spanning trees with $n \leq 3$, result in a single graph equivalence class and do not need rewiring. Given an arbitrary spanning tree with $n > 3$, the transition to the Linear graph requires a degree lowering operation. 
 
\begin{figure}
	\centering
	\subfloat[]{
	\begin{tikzpicture} 
		\tikzstyle{every node} = [draw, shape=circle, minimum size=0.9cm]
		\node(v1) at (0:0) {$H$};
		\node(v2) at (180:1.5) {$M$};
		\node(v3) at (270:1.5) {$T$};
		\node(v4) at (0:1.5) {$\ldots$};
		\node(v5) at (-3, 0) {$\ldots$};
		\node(v6) at (90:1.5) {$\ldots$};

		\draw (v1) -- node[strike out, draw, -, minimum size=0.2cm]{}(v2);
		\draw (v1) -- (v3) node[fill=none, draw=none, midway, right] {$e$};
		\draw[dashed] (v2) to[out=-90,in=180] (v3);
		\draw (v1) -- (v2);
		\draw (v1) -- (v3);
		\draw (v1) -- (v4);
		\draw (v5) -- (v2);
		\draw (v1) -- (v6);
	\end{tikzpicture}
	}\\ \bigskip
	\subfloat[]{
	\begin{tikzpicture}
		\tikzstyle{every node} = [draw, shape=circle, minimum size=0.9cm]
		\node(1) at (1.5,0) {$M$};
		\node(2) at (3,0) {$T$};
		\node(3) at (4.5,0) {$H$};
		\node(4) at (5.25,1.3) {$\ldots$};
		\node(5) at (0, 0) {$\ldots$};
		\node(6) at (5.25,-1.3) {$\ldots$};

		\draw (2) -- (3) node[fill=none, draw=none, midway, above] {$e$};
		\draw (5) -- (1);
		\draw (1) -- (2);
		\draw (2) -- (3);
		\draw (3) -- (4);
		\draw (6) -- (3);	
	\end{tikzpicture} 
	}
	\caption{This transition lowers the degree of $H$. Nodes adjacent to the hub marked $H$ are rewired and become part of the Linear subgraph containing $T$. Repeated application of this transition to all nodes with $k > 2$ transforms the graph into the Linear graph.}
	\label{fig:degree_lowering}
\end{figure}
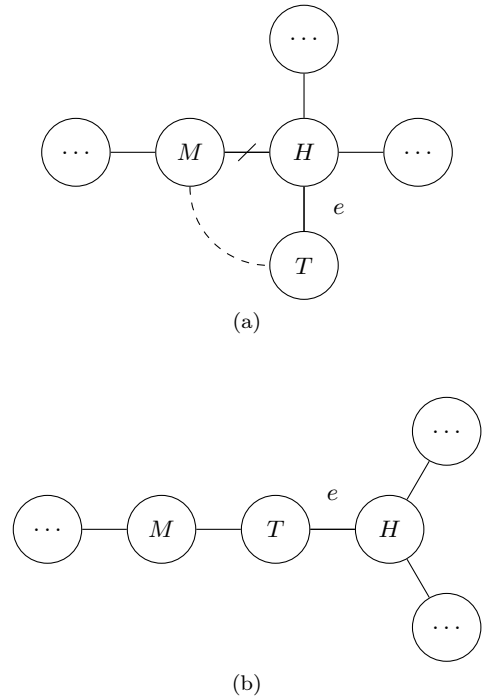

This transition is considered here on a Star graph. An edge incident to a node with $k(v_i)>2$ is chosen with probability $p_e = 1/(n-1)$.  The central node has degree, $k(v_i) = (n-1)$ and all other nodes have degree, $k(v_i) = 1$. Using our algorithm, the center and one of the radial vertices are labeled as $H$  and $T$ respectively. A second vertex adjacent to $H$ is chosen as the $M$ vertex, with probability $p=1/(k(H)-1)$. By performing the rewiring in Fig.~\ref{fig:degree_lowering}, the degree of $H$ is lowered. The radial nodes can be repeatedly chosen as the $T$ node. This lowers the degree of the central node until the max degree in the graph is two. 

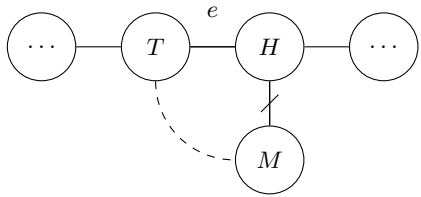
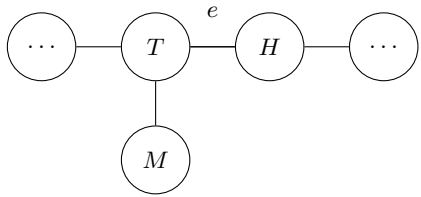
\begin{figure}[H]
	\centering
	\subfloat[A node $M$ is rewired away from a hub $H$. This shifts $M$ towards the end of the Linear subgraph attached to $T$. ]{
	\begin{tikzpicture} 
		\tikzstyle{every node} = [draw, shape=circle, minimum size=0.9cm]
		\node(0) at (0,0) {$\ldots$};
		\node(1) at (1.5,0) {$T$};
		\node(2) at (3,0) {$H$};
		\node(3) at (4.5,0) {$\ldots$};
		\node(4) at (3,-1.5) {$M$};

		\draw[dashed] (1) to[out=-90,in=180] (4);
		\draw (0) -- (1);
		\draw (1) -- (2) node[fill=none, draw=none, midway, above] {$e$};
		\draw (2) -- node[strike out, draw, -, minimum size=0.2cm]{}(4);
		\draw (1) -- (2);
		\draw (2) -- (3);
		\draw (2) -- (4);
	\end{tikzpicture}
	}\\ \bigskip
	\subfloat[$M$ is shifted along the subgraph by a distance of one edge. Repeating this operation decreases the degree of any hubs and places the $M$-type nodes at the end of the Linear graph.]{
	\begin{tikzpicture}
		\tikzstyle{every node} = [draw, shape=circle, minimum size=0.9cm]
		\node(v0) at (0,0) {$\ldots$};		
		\node(v1) at (1.5,0) {$T$};
		\node(v2) at (3,0) {$H$};
		\node(v3) at (4.5,0) {$\ldots$};
		\node(v4) at (1.5,-1.5) {$M$};

		\draw (v0) -- (v1);
		\draw (v1) -- (v2) node[fill=none, draw=none, midway, above] {$e$};
		\draw (v1) -- (v2);
		\draw (v2) -- (v3);
		\draw (v1) -- (v4);

	\end{tikzpicture}
	}
	\caption{Translating a node, $M$, to the end of a Linear subtree.}
	\label{fig:translating_subtree}
\end{figure}

More generally, we can choose to rewire edges where $k(T) > 1$ also. The rewiring of the hub node will attach $M$ to $T$, even though $T$ is not a terminal vertex and this subgraph is not Linear. The rewiring is repeated and $M$ can either be translated down the $T$ subgraph as in Fig.~\ref{fig:translating_subtree} or the $T$ subgraph can be can be rewired onto $M$. This transforms this segment of the subgraph into the Linear graph. This is repeated for all $H$ nodes with degree greater than two.
Therefore, given any tree we can decrease the degree of any node with $k(v_i)>2$ using two different methods. This can be repeated for all nodes with degree $k(v_i)>2$, until we get the Linear graph.
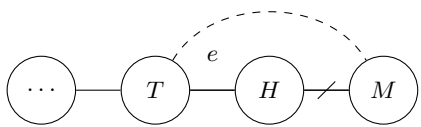
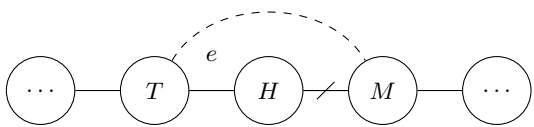
\begin{figure}[H]

	\centering
	\subfloat[Rewiring at the end-node of a graph.]{
	\begin{tikzpicture} \label{graph:endpoint_rewire}
		\tikzstyle{every node} = [draw, shape=circle, minimum size=0.9cm]
		\node(1) at (1.5,0) {$\ldots$};
		\node(2) at (3,0) {$T$};
		\node(3) at (4.5,0) {$H$};
		\node(4) at (6,0) {$M$};

		\draw[dashed] (2) to[out=60,in=120] (4);
		\draw (2) -- (3) node[fill=none, draw=none, above, midway] {$e$};
		\draw (3) -- node[strike out, draw, -, minimum size=0.2cm]{}(4);
		\draw (1) -- (2);
		\draw (2) -- (3);
		\draw (3) -- (4);
	\end{tikzpicture}
	}\\ \bigskip
	\subfloat[Rewiring at a mid-point node of a graph.]{
	\centering
	\begin{tikzpicture} \label{graph:midpoint_rewire}
		\tikzstyle{every node} = [draw, shape=circle, minimum size=0.9cm]
		\node(1) at (0,0) {$\ldots$};
		\node(2) at (1.5,0) {$T$};
		\node(3) at (3,0) {$H$};
		\node(4) at (4.5,0) {$M$};
		\node(5) at (6,0) {$\ldots$};

		\draw[dashed] (2) to[out=60,in=120] (4);
		\draw (2) -- (3) node[fill=none, draw=none, above, midway] {$e$};
		\draw (3) -- node[strike out, draw, -, minimum size=0.2cm]{}(4);
		\draw (1) -- (2);
		\draw (2) -- (3);
		\draw (3) -- (4);
		\draw (4) -- (5);
	\end{tikzpicture}
	}
\caption{Increasing the degree of a specific node to create a hub. The operation in Fig.~\ref{graph:midpoint_rewire} can be repeated to increase the degree of the node marked $T$ here.}
\label{fig:degree_increasing}
\end{figure}
In order to transition from the Linear graph back to any other graph in the ensemble, we must be able to increase the degree of any node in any position. To demonstrate this, we create a Star graph from the Linear graph. It is now necessary to have a degree increasing operation in addition to the translation operation. A degree increasing operation can be performed at any position in a subgraph. Fig.~\ref{fig:degree_increasing} illustrates a degree increasing rewire at the endpoint and at a position towards the middle of a subgraph. By performing a degree increasing operation at the endpoint of a graph and translating the resulting branch down the graph, we can create any spanning tree including the star graph.

So far we have demonstrated rewiring operations which can;
\begin{itemize}
\item{lower the degree of a node,}
\item{increase the degree of a node near the endpoint of graph,}
\item{increase the degree of a node in middle of graph and}
\item{translate a node of any degree to any position in tree.}
\end{itemize}
Since these operations can be used to transition from any graph to the Linear graph and back, the Markov chain is irreducible. It should be noted that all of these transitions are performed using the same rewiring technique and the result of a rewire differs only depending on the neighbor structure of the edge chosen to rewire around.

\subsubsection{Aperiodicity} 
The period of the state $x$, $o_x$, is the greatest common divisor of the number of transitions $n$, such that $p_{xx}^{(n)} > 0 \hspace{0.2cm}\forall \hspace{0.2cm} n$. All states of an irreducible Markov chain have the same periodicity and any state $x$ is aperiodic if $o_x = 1$. To prove that our Markov chain is aperiodic, we will demonstrate that a transition between the Linear and Star graphs exists in which $o_x=1$. 

Starting with the Linear graph, we will show that there exists two consecutive numbers, $t$ and $s$, that satisfy the $n$-step transition back to state $x$ such that $p_{xx}^{(n)}>0$. Since two consecutive natural numbers have a greatest common divisor of one, the existence of $t$ and $s$ is sufficient to prove that our Markov chain is aperiodic.

\begin{figure}[H]
	\centering
	\subfloat[]{
	\begin{tikzpicture} 
		\tikzstyle{every node} = [draw, shape=circle, minimum size=0.9cm]
		\node(1) at (0:0) {$H$};
		\node(2) at (72:1.5) {$M$};
		\node(3) at (144:1.5) {$T$};
		\node(4) at (216:1.5) {};
		\node(5) at (288:1.5) {};
		\node(6) at (0:1.5) {};

		\draw[dashed] (2) to[out=90,in=90] (3);
		\draw (1) -- (3) node[fill=none, draw=none, above, midway] {$e$};
		\draw (1) -- node[strike out, draw, -,minimum size =0.2cm]{}(2);
		\draw (1) -- (2);
		\draw (1) -- (3);
		\draw (1) -- (4);
		\draw (1) -- (5);
		\draw (1) -- (6);
	\end{tikzpicture}
	}\\ \bigskip
	\subfloat[]{
	\begin{tikzpicture} 
		\tikzstyle{every node} = [draw, shape=circle, minimum size=0.9cm]
		\node(1) at (0,0) {};
		\node(2) at (1.5,0) {$M$};
		\node(3) at (3,0) {$H$};
		\node(4) at (3,1.5) {$T$};
		\node(5) at (4.5,0) {};
		\node(6) at (3,-1.5) {};

		\draw[dashed] (2) to[out=90,in=180] (4);
		\draw (4) -- (3) node[fill=none, draw=none, right, midway] {$e$};
		\draw (2) -- node[strike out, draw, -,minimum size =0.2cm]{}(3);
		\draw (1) -- (2);
		\draw (2) -- (3);
		\draw (3) -- (4);
		\draw (3) -- (5);
		\draw (3) -- (6);
	\end{tikzpicture}
	} \\ \bigskip
	\subfloat[]{
	\centering
	\begin{tikzpicture} 
		\tikzstyle{every node} = [draw, shape=circle, minimum size=0.9cm]
		\node(1) at (0,0) {};
		\node(2) at (1.5,0) {};
		\node(3) at (3,0) {$M$};
		\node(4) at (4.5,0) {$H$};
		\node(5) at (5.25,1.3) {$T$};
		\node(6) at (5.25,-1.3) {};

		\draw[dashed] (3) to[out=90,in=180] (5);
		\draw (4) -- (5) node[fill=none, draw=none, right, midway] {$e$};
		\draw (3) -- node[strike out, draw, -,minimum size =0.2cm]{}(4);
		\draw (1) -- (2);
		\draw (2) -- (3);
		\draw (3) -- (4);
		\draw (4) -- (5);
		\draw (4) -- (6);
	\end{tikzpicture}
	} \\ \bigskip
	\subfloat[]{
	\centering
	\begin{tikzpicture} 
		\tikzstyle{every node} = [draw, shape=circle, minimum size=0.9cm]
		\node(1) at (0,0) {};
		\node(2) at (1.5,0) {};
		\node(3) at (3,0) {};
		\node(4) at (4.5,0) {};
		\node(5) at (6.0,0) {};
		\node(6) at (7.5,0) {};

		\draw (1) -- (2);
		\draw (2) -- (3);
		\draw (3) -- (4);
		\draw (4) -- (5);
		\draw (5) -- (6);
	\end{tikzpicture}
	}
\caption{Transitioning from a Star to a Linear graph in $n-3$ operations.}
\label{fig:star_to_line}
\end{figure}
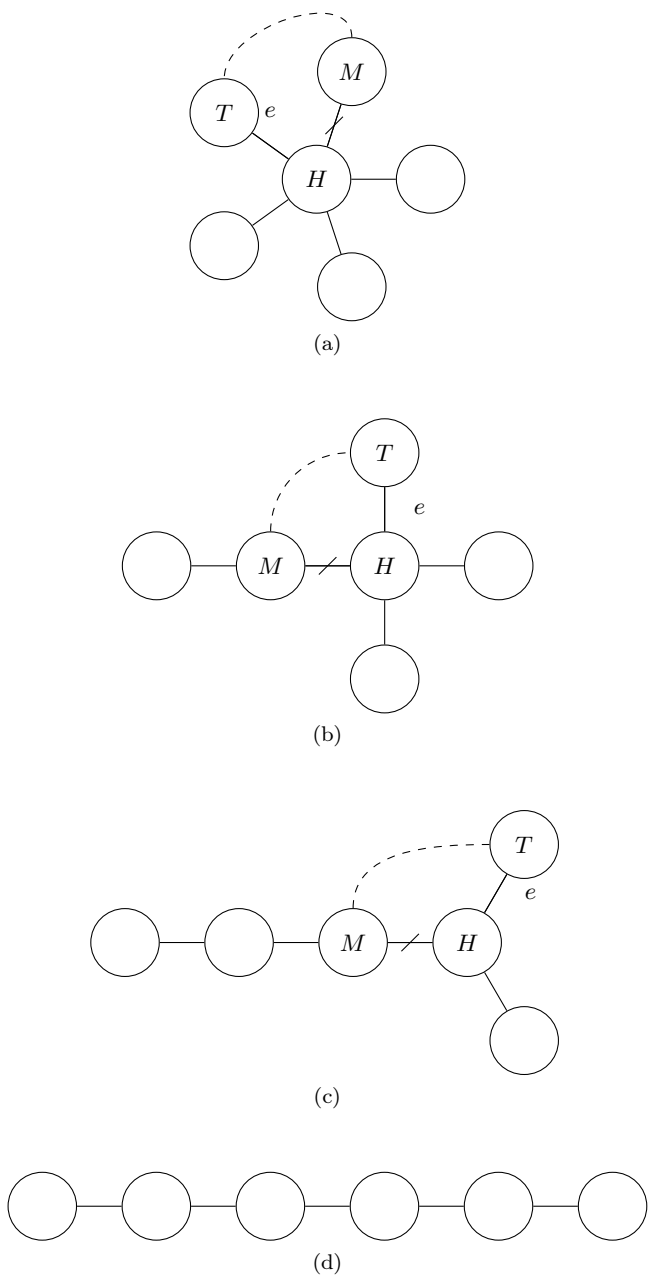

\begin{figure}[H]
	\centering
	\subfloat[]{
	\centering
	\begin{tikzpicture} 
		\tikzstyle{every node} = [draw, shape=circle, minimum size=0.9cm]
		\node(1) at (0,0) {};
		\node(2) at (1.5,0) {};
		\node(3) at (3,0) {};
		\node(4) at (4.5,0) {$T$};
		\node(5) at (6.0,0) {$H$};
		\node(6) at (7.5,0) {$M$};

		\draw[dashed] (4) to[out=60,in=120] (6);
		\draw (4) -- (5) node[fill=none, draw=none, above, midway] {$e$};
		\draw (5) -- node[strike out, draw, -,minimum size =0.2cm]{}(6);
		\draw (1) -- (2);
		\draw (2) -- (3);
		\draw (3) -- (4);
		\draw (4) -- (5);
		\draw (5) -- (6);
	\end{tikzpicture}
	}\\ \bigskip
	\subfloat[]{
	\centering
	\begin{tikzpicture} 
		\tikzstyle{every node} = [draw, shape=circle, minimum size=0.9cm]
		\node(1) at (0,0) {};
		\node(2) at (1.5,0) {$M$};
		\node(3) at (3,0) {$H$};
		\node(4) at (4.5,0) {$T$};
		\node(5) at (5.25,1.3) {};
		\node(6) at (5.25,-1.3) {};

		\draw[dashed] (2) to[out=60,in=120] (4);
		\draw (3) -- (4) node[fill=none, draw=none, above, midway] {$e$};
		\draw (2) -- node[strike out, draw, -,minimum size =0.2cm]{}(3);
		\draw (1) -- (2);
		\draw (2) -- (3);
		\draw (3) -- (4);
		\draw (4) -- (5);
		\draw (4) -- (6);
	\end{tikzpicture}
	}\\ \bigskip
	\subfloat[]{
	\begin{tikzpicture} 
		\tikzstyle{every node} = [draw, shape=circle, minimum size=0.9cm]
		\node(1) at (0,0) {$M$};
		\node(2) at (1.5,0) {$H$};
		\node(3) at (3,0) {$T$};
		\node(4) at (3,1.5) {};
		\node(5) at (4.5,0) {};
		\node(6) at (3,-1.5) {};

		\draw[dashed] (1) to[out=60,in=120] (3);
		\draw (2) -- (3) node[fill=none, draw=none, above, midway] {$e$};
		\draw (1) -- node[strike out, draw, -,minimum size =0.2cm]{}(2);
		\draw (1) -- (2);
		\draw (2) -- (3);
		\draw (3) -- (4);
		\draw (3) -- (5);
		\draw (3) -- (6);
	\end{tikzpicture}
	}\\ \bigskip
	\subfloat[]{
	\begin{tikzpicture} 
		\tikzstyle{every node} = [draw, shape=circle, minimum size=0.9cm]
		\node(1) at (0:0) {};
		\node(2) at (72:1.5) {};
		\node(3) at (144:1.5) {};
		\node(4) at (216:1.5) {};
		\node(5) at (288:1.5) {};
		\node(6) at (0:1.5) {};

		\draw (1) -- (2);
		\draw (1) -- (3);
		\draw (1) -- (4);
		\draw (1) -- (5);
		\draw (1) -- (6);
	\end{tikzpicture}
	}	
\caption{Transitioning from a Linear to a Star graph in $n-3$ operations.}
\label{fig:line_to_star}
\end{figure}
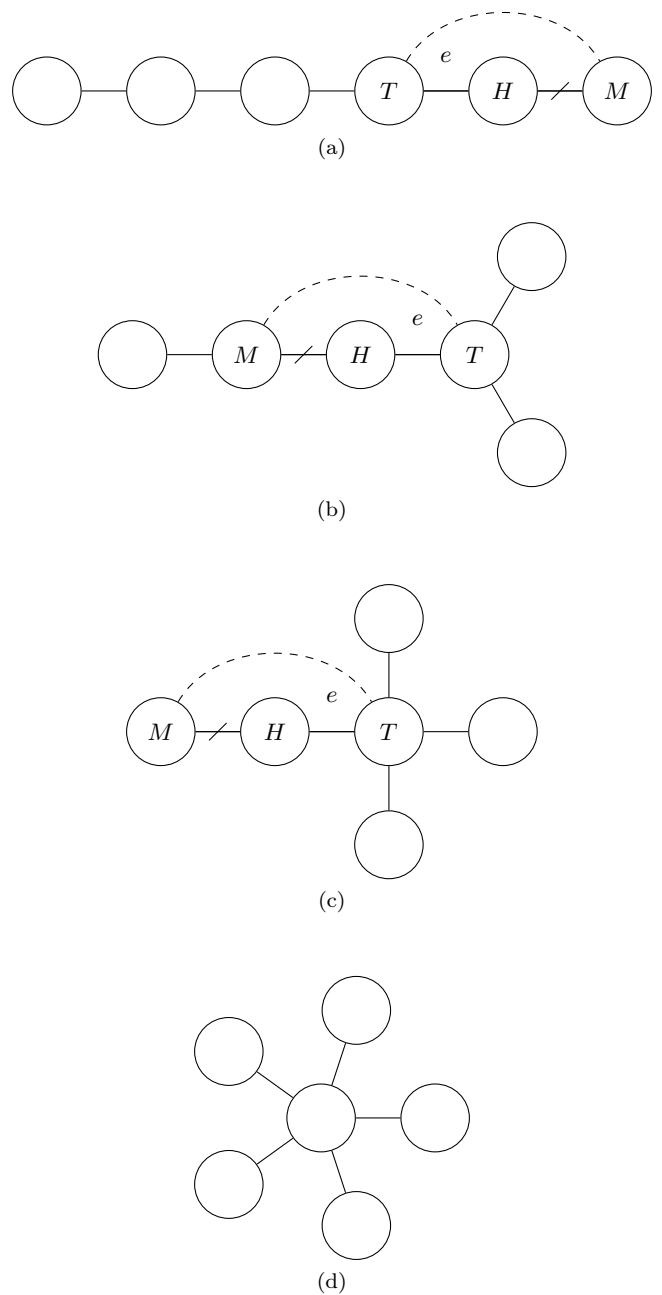

A Star graph containing $n$ nodes has a central hub node with degree of $n-1$. It is possible to transition from a Star graph to a Linear graph in $q=n-3$ operations by rewiring an edge incident to the central node $n-3$ times. This is shown in Fig.~\ref{fig:star_to_line}.
The inverse transformation from the Linear graph back to the Star using $q=n-3$ operations also exists and is shown in Fig.~\ref{fig:line_to_star}.

This transformation can be considered using the $n$-step transition probabilities: $p^{q}_{\mathrm{Star}\rightarrow\mathrm{Linear}}>0$ and $p^{q}_{\mathrm{Linear}\rightarrow\mathrm{Star}}>0$. Define $t$ so that 
\begin{align*}
p^{t}_{StarStar} = \left(p^{q}_{\mathrm{Star}\rightarrow\mathrm{Linear}} \cdot p^{q}_{\mathrm{Linear}\rightarrow\mathrm{Star}}\right) > 0.
\end{align*}
The number of rewiring operations necessary to perform this transformation is $t = 2q = 2(n-3)$ is even for all graph sizes. 

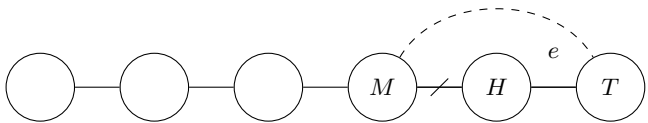
\begin{figure}[H]
	\centering
	\begin{tikzpicture} 
		\tikzstyle{every node} = [draw, shape=circle, minimum size=0.9cm]
		\node(1) at (0,0) {};
		\node(2) at (1.5,0) {};
		\node(3) at (3,0) {};
		\node(4) at (4.5,0) {$M$};
		\node(5) at (6.0,0) {$H$};
		\node(6) at (7.5,0) {$T$};

		\draw[dashed] (4) to[out=60,in=120] (6);
		\draw (5) -- (6) node[fill=none, draw=none, above, midway] {$e$};
		\draw (4) -- node[strike out, draw, -,minimum size =0.2cm]{}(5);
		\draw (1) -- (2);
		\draw (2) -- (3);
		\draw (3) -- (4);
		\draw (4) -- (5);
		\draw (5) -- (6);
	\end{tikzpicture}

\caption{Graph rewiring which maps the Linear graph back to itself.}
\label{fig:line_to_star_slower}
\end{figure}

It is possible to perform the round-trip from Star to Star in many other ways. We define the value of $s$ by taking one of these other paths in state space. Perform the $q$ operations from the Star to Linear graph just like above. However, we perform a different return sequence of transitions. Either perform the trivial rewire in Fig.~\ref{fig:line_to_star_slower}, or a subgraph translation operation as in Fig.~\ref{fig:translating_subtree} in addition to the $q$ necessary operations to transition from the Linear to the Star graphs. Overall, $r=q+1$ operations are used to transition from the Linear graph back to the Star in this manner. Therefore, $p^{s}_{xx}>0$ for $s = r+q = 2q+1$, which must be odd. The period of the Star graph is then given by the greatest common denominator of $t$ and $s$. The period is one, since $t$ and $s$ are consecutive integers and must have greatest common divisor of one. Therefore our Markov chain is aperiodic.  
\subsubsection{Positive Recurrence}
Finally, since any irreducible finite state Markov chain is positive recurrent, it follows that our algorithm satisfies the conditions necessary for ergodicity. Therefore, there exists a unique stationary distribution $\pi$ and is given by
\begin{align*}
\pi_j = \lim_{n \to \infty} p^{(n)}_{ij}.
\end{align*}
The stationary distribution is also universal, since it does not depend on the initial state of the system.

\subsection{Graph Diameter Observable}\label{sec:graph_diameter}
Having proven the ergodicity of our algorithm, we can use it to estimate the ensemble average of graph observables. The graph diameter, $d$, of a spanning tree is defined as the longest path between any pair of vertices in the graph. This observable depends on the overall graph structure and provides a measure of the maximum distance that information needs to travel in a wireless-telecommunications tree network. 

Undirected spanning trees have two special properties: there exists only one path between any pair of nodes and any node can be chosen as a root of the graph. In order to calculate the graph diameter, we perform a breadth-first search (BFS) to find the furthest node from a randomly chosen root node, $r$. The BFS algorithm searches over the immediate neighbors of $r$ first. The neighbors are put in a queue to have their distance inspected in turn. Each of these nodes incident to $r$ will have a distance of one. Each subsequent level of neighbors will have a distance of the current node plus one. Eventually, all nodes will have been visited and the furthest node $v_1$ from $r$ is recorded. 
\begin{figure}[H]
	\centering
	\begin{tikzpicture} 
	    \definecolor{tblue}{rgb}{0.0,0.565,0.706}
		\tikzstyle{every node} = [draw, shape=circle, minimum size=0.9cm]
		\node[fill = tblue](-1) at (-.75, 1.3) {};
		\node(0) at (-.75, -1.3) {};
		\node(1) at (0,0) {};
		\node(2) at (1.5,0) {};
		\node(3) at (3,0) {};
		\node(4) at (4.5,0) {};
		\node(5) at (5.25, 1.3) {};
		\node(6) at (5.25, -1.3) {};
		\node[fill = tblue](7) at (6.0,0) {};
		\node(9) at (1.5, 1.5) {};
		\node(10) at (1.5, -1.5) {};

		\draw[tblue,line width=2pt] (-1) -- (1);
		\draw (0) -- (1);
		\draw[tblue,line width=2pt] (1) -- (2);
		\draw[tblue,line width=2pt] (2) -- (3);
		\draw[tblue,line width=2pt] (3) -- (4);
		\draw[tblue,line width=2pt] (4) -- (7);
		\draw (4) -- (6);
		\draw (4) -- (5);
		\draw (2) -- (9);
		\draw (2) -- (10);
	\end{tikzpicture}
	
\caption{The shaded nodes and edges mark the longest path. The diameter is the length of this longest path. The diameter depends on the structure of all nodes and edges in the graph. To calculate the diameter, the distance of all respective pairs of nodes must be tested. The diameter of a tree is calculated by performing a Breadth-First Search from any node, to identify one of the endpoints, $v_1$. By performing a second Breadth-First Search with $v_1$ as the root node, the second endpoint $v_2$ of the longest path is found to be the node furthest from $v_1$. The distance between $v_1$ and $v_2$ is the graph diameter. In this graph, the longest path is not unique, but this does not change the diameter.}
\label{fig:bfs_diameter}
\end{figure}
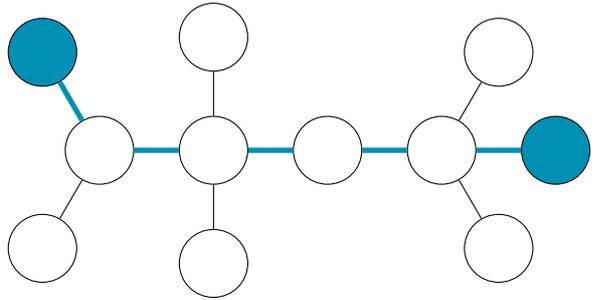

Since there exists only one path between any pair of nodes, the furthest node $v_1$ from the root $r$ must be one of the endpoints of the longest path. There may be many nodes which are equidistant from $r$, in which case any one of these is a valid choice for $v_1$. To calculate the diameter, a second BFS is rooted at the node $v_1$, since the furthest node from an endpoint of the longest path is the other endpoint. The second endpoint $v_2$ is identified and the distance between them is returned by the BFS. In a spanning tree, the largest graph distance is the diameter.

Using our rewiring algorithm, we can estimate the expected value of the graph diameter over the graph ensemble,
\begin{align*}
\bar{d} \approx \langle d \rangle = \sum_{g_i \in G} d(g_i) \pi_{g_i}.
\end{align*}
A very similar problem has already been addressed asymptotically in Ref.~\cite{Szekeres:67}. In that paper, the authors focus on the height of a spanning tree, $g$, with ordered vertices, $P_1, P_2, \ldots, P_n$. The height of a tree, $h_{P_1}(g)$, is defined as the length of the longest path in $g$ from an arbitrarily chosen root node $P_1$. An expression for the asymptotic distribution of the number of spanning trees with $n$ vertices having exactly height, $k$, is found using a recursion relation. In the limit where $n$ and $k$ are large, the authors go on to calculate an expectation value of the height of trees starting from $P_1$ over the set of the $n^{n-2}$ spanning trees with $n$ nodes. It should be noted, the height of a tree from a node $P_1$ is bounded by the diameter of that tree, since $P_1$ is not guaranteed to be an endpoint of the longest path,
\begin{align}
\frac{1}{2}d(g) \leq \min_{i}h_{P_i}(g) \leq h_{P_i} (g) \leq \max_{i} h_{P_i}(g) \leq d(g). \label{eq: tree_height_bound}
\end{align}
The expectation value of the height of trees with root node $P_1$ as a function of the number of nodes $|V|$ given in Ref.~\cite{Szekeres:67} using the asymptotic probability distribution described above is
\begin{align}\label{eq:expectation_height}
\langle h_{P_1}\rangle \approx \sqrt{2 \pi |V|} = 2.50663\sqrt{|V|}.
\end{align}
In the next section, we will show our numerical estimate of the mean diameter using our algorithm over three orders of magnitude of graph size. The line of best fit of graph diameter versus graph size $\lvert V\rvert$ is calculated using the non-linear least squares method and compared with the asymptotic expectation value of Eq.~\ref{eq:expectation_height}. Errors quoted for the parameters are estimated by the square root of the diagonals of the parameter covariance matrix.

\section{Results and Discussion \label{sec:results}} 

Finally, in this section we present the results of our Monte Carlo simulations. Our results demonstrate that our algorithm samples from the graph distribution efficiently. For small spanning trees, we compare our Monte Carlo estimates of the graph distribution, ($\bar{\pi}$), with the exact distribution. We also compare the estimated mean of the graph diameter for small graphs with the expectation value. These estimates of the mean graph diameter and their statistical errors are compared for the $K_4$, $K_5$, $K_6$ and $K_7$ ensembles. The integrated autocorrelation times of the graph diameter are shown to be under control as we increase the graph size and finally, the asymptotic behavior of the graph diameter is discussed as the graph size increases.

\subsection{Graph Distribution}\label{rewiring:graph_dist}

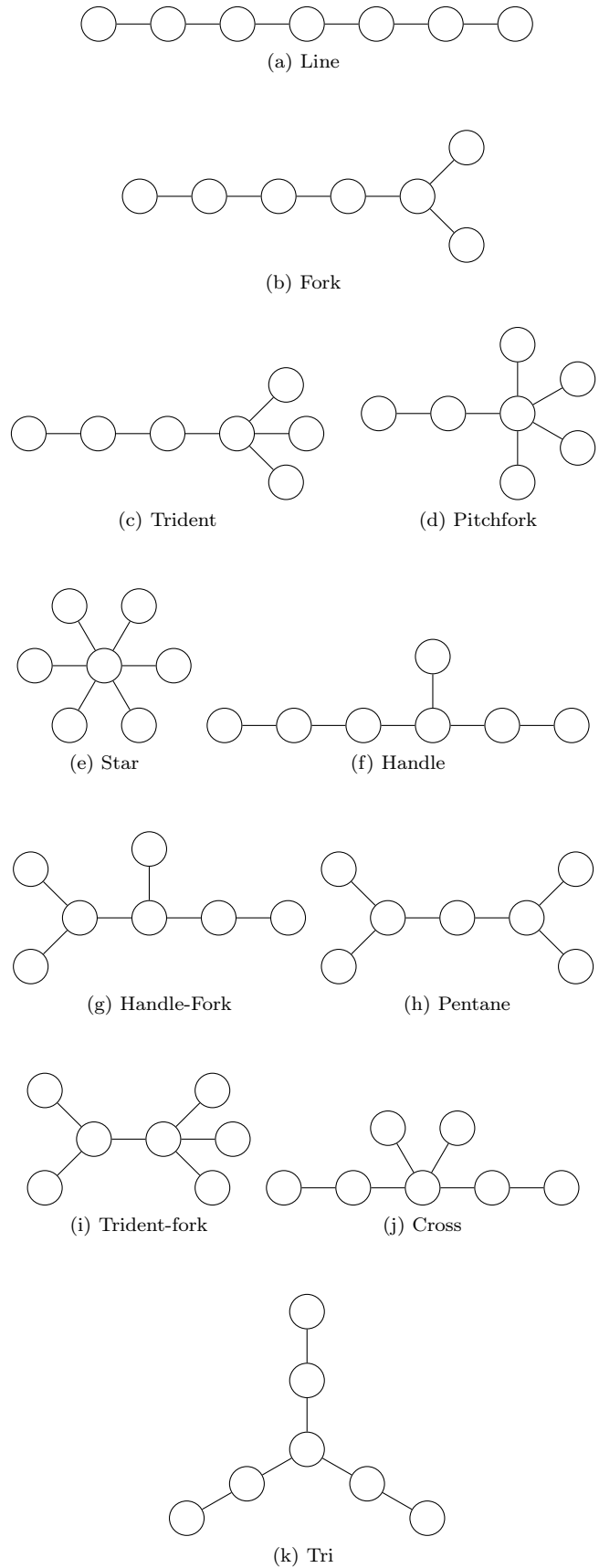
\begin{figure}[H] 
	\centering
	\subfloat[Line]{
	\begin{tikzpicture} \label{graph:n7_line}
		\tikzstyle{every node} = [draw, shape=circle, minimum size = 0.5cm]
		\node(v1) at (0,0) {};
		\node(v2) at (1,0) {};
		\node(v3) at (2,0) {};
		\node(v4) at (3,0) {};
		\node(v5) at (4,0) {};
		\node(v6) at (5,0) {};
		\node(v7) at (6,0) {};

		\draw (v1) -- (v2);
		\draw (v2) -- (v3);
		\draw (v3) -- (v4);
		\draw (v4) -- (v5);
		\draw (v5) -- (v6);
		\draw (v6) -- (v7);
	\end{tikzpicture}}\\ \bigskip
	\subfloat[Fork]{
		\begin{tikzpicture} \label{graph:n7_fork}
			\tikzstyle{every node} = [draw, shape=circle, minimum size = 0.5cm]
			\node(v1) at (0,0) {};
			\node(v2) at (1,0) {};
			\node(v3) at (2,0) {};
			\node(v4) at (3,0) {};
			\node(v5) at (4,0) {};
			\node(v6) at (4.707,0.707) {};
			\node(v7) at (4.707,-0.707) {};
			\draw (v1) -- (v2);
			\draw (v2) -- (v3);
			\draw (v3) -- (v4);
			\draw (v4) -- (v5);
			\draw (v5) -- (v6);
			\draw (v5) -- (v7);
		\end{tikzpicture}
	} \\
	\bigskip
	\subfloat[Trident]{
		\begin{tikzpicture} \label{graph:n7_trident}
			\tikzstyle{every node} = [draw, shape=circle, minimum size = 0.5cm]
			\node(v1) at (0,0) {};
			\node(v2) at (1,0) {};
			\node(v3) at (2,0) {};
			\node(v4) at (3,0) {};
			\node(v5) at (3.707,0.707) {};
			\node(v6) at (4,0) {};
			\node(v7) at (3.707,-0.707) {};
			\draw (v1) -- (v2);
			\draw (v2) -- (v3);
			\draw (v3) -- (v4);
			\draw (v4) -- (v5);
			\draw (v4) -- (v6);
			\draw (v4) -- (v7);
		\end{tikzpicture}
	} \hspace{0.2cm}
	\subfloat[Pitchfork]{
		\begin{tikzpicture} \label{graph:n7_pitchfork}
			\tikzstyle{every node} = [draw, shape=circle, minimum size = 0.5cm]
			\node(v1) at (0,0) {};
			\node(v2) at (1,0) {};
			\node(v3) at (2,0) {};
			\node(v4) at (2,1) {};
			\node(v5) at (2.866,0.5) {};
			\node(v6) at (2.866,-0.5) {};
			\node(v7) at (2,-1) {};
			\draw (v1) -- (v2);
			\draw (v2) -- (v3);
			\draw (v3) -- (v4);
			\draw (v3) -- (v5);
			\draw (v3) -- (v6);
			\draw (v3) -- (v7);
		\end{tikzpicture}
	} \\
	\bigskip
	\subfloat[Star]{
		\begin{tikzpicture} \label{graph:n7_star}
			\tikzstyle{every node} = [draw, shape=circle, minimum size = 0.5cm]
			\node(v1) at (0:0) {};
			\node(v2) at (0:1) {};
			\node(v3) at (60:1) {};
			\node(v4) at (120:1) {};
			\node(v5) at (180:1) {};
			\node(v6) at (240:1) {};
			\node(v7) at (300:1) {};
			\draw (v1) -- (v2);
			\draw (v1) -- (v3);
			\draw (v1) -- (v4);
			\draw (v1) -- (v5);
			\draw (v1) -- (v6);
			\draw (v1) -- (v7);
		\end{tikzpicture}
	} 
	\subfloat[Handle]{
		\begin{tikzpicture} \label{graph:n7_handle}
			\tikzstyle{every node} = [draw, shape=circle, minimum size = 0.5cm]
			\node(v1) at (1,0) {};
			\node(v2) at (2,0) {};
			\node(v3) at (3,0) {};
			\node(v4) at (4,0) {};
			\node(v5) at (5,0) {};
			\node(v6) at (6,0) {};
			\node(v7) at (4,1) {};
			\draw (v1) -- (v2);
			\draw (v2) -- (v3);
			\draw (v3) -- (v4);
			\draw (v4) -- (v5);
			\draw (v5) -- (v6);
			\draw (v4) -- (v7);
		\end{tikzpicture}
	}\\
	\bigskip
	\subfloat[Handle-Fork]{
		\begin{tikzpicture} \label{graph:n7_handlefork}
			\tikzstyle{every node} = [draw, shape=circle, minimum size = 0.5cm]
			\node(v1) at (0,0.707) {};
			\node(v2) at (0,-0.707) {};
			\node(v3) at (0.707,0) {};
			\node(v4) at (1.707,0) {};
			\node(v5) at (2.707,0) {};
			\node(v6) at (3.707,0) {};
			\node(v7) at (1.707,1) {};
			\draw (v1) -- (v3);
			\draw (v2) -- (v3);
			\draw (v3) -- (v4);
			\draw (v4) -- (v5);
			\draw (v5) -- (v6);
			\draw (v4) -- (v7);
		\end{tikzpicture}
	} 
		\subfloat[Pentane]{
		\begin{tikzpicture} \label{graph:n7_pentane}
			\tikzstyle{every node} = [draw, shape=circle, minimum size = 0.5cm]
			\node(v1) at (0,0.707) {};
			\node(v2) at (0,-0.707) {};
			\node(v3) at (0.707,0) {};
			\node(v4) at (1.707,0) {};
			\node(v5) at (2.707,0) {};
			\node(v6) at (3.414,0.707) {};
			\node(v7) at (3.414,-0.707) {};
			\draw (v1) -- (v3);
			\draw (v2) -- (v3);
			\draw (v3) -- (v4);
			\draw (v4) -- (v5);
			\draw (v5) -- (v6);			
			\draw (v5) -- (v7);
		\end{tikzpicture}
	} \\ \bigskip
	\subfloat[Trident-fork]{
		\begin{tikzpicture} \label{graph:n7_trifork}
			\tikzstyle{every node} = [draw, shape=circle, minimum size = 0.5cm]
			\node(v1) at (0,0.707) {};
			\node(v2) at (0,-0.707) {};
			\node(v3) at (0.707,0) {};
			\node(v4) at (1.707,0) {};
			\node(v5) at (2.414,0.707) {};
			\node(v6) at (2.707,0) {};
			\node(v7) at (2.414,-0.707) {};
			\draw (v1) -- (v3);
			\draw (v2) -- (v3);
			\draw (v3) -- (v4);
			\draw (v4) -- (v5);
			\draw (v4) -- (v6);
			\draw (v4) -- (v7);
		\end{tikzpicture}
	} 
		\subfloat[Cross]{
		\begin{tikzpicture} \label{graph:n7_2handle}
			\tikzstyle{every node} = [draw, shape=circle, minimum size = 0.5cm]
			\node(v1) at (0,0) {};
			\node(v2) at (1,0) {};
			\node(v3) at (2,0) {};
			\node(v4) at (3,0) {};
			\node(v5) at (4,0) {};
			\node(v6) at (1.5,0.866) {};
			\node(v7) at (2.5,0.866) {};
			\draw (v1) -- (v2);
			\draw (v2) -- (v3);
			\draw (v3) -- (v4);
			\draw (v4) -- (v5);
			\draw (v3) -- (v6);
			\draw (v3) -- (v7);
		\end{tikzpicture}
	}\\ \bigskip
	\subfloat[Tri]{
	\begin{tikzpicture}[rotate = -90] \label{graph:n7_Tri}
		\tikzstyle{every node} = [draw, shape=circle, minimum size = 0.5cm]
		\node(v1) at (0:0) {};
		\node(v2) at (180:1) {};
		\node(v3) at (180:2) {};
		\node(v4) at (300:1) {};
		\node(v5) at (300:2) {};
		\node(v6) at (60:1) {};
		\node(v7) at (60:2) {};

		\draw (v1) -- (v2);
		\draw (v2) -- (v3);
		\draw (v1) -- (v4);
		\draw (v4) -- (v5);
		\draw (v1) -- (v6);
		\draw (v6) -- (v7);
	\end{tikzpicture}}

	\caption{Non-isomorphic spanning trees of $K_{7}$. All spanning trees of $K_7$ are isomorphic to one of these graphs.}\label{graph:K_7_spanning_trees}
\end{figure}

Here we present and compare the Monte Carlo estimated probability distribution, $\bar{\pi}$, with the exactly known probability distributions of the $K_7$ ensemble. When working with spanning trees containing few nodes, the spanning tree isomorphism classes can be identified by hand, as in Fig.~\ref{graph:K_7_spanning_trees}. For ensembles of graphs larger than $K_7$, the high amount of possible branching makes it very difficult to identify all of the graph isomorphism classes. The estimated probability distribution, $\bar{\pi}_{g_i}$, for the ensemble of spanning trees of $K_7$ is shown in Tab~\ref{table:k7_results}. 

\begin{table*}[t]
	\centering	
	\begin{ruledtabular}	
	\begin{tabular}{l c c c c c} 
	Tree $(n^{n-2}=16807)$ & $\lvert \mathrm{Aut}(g_i) \rvert$ & $l_{g_i}$ & $\pi_{g_{i}}$ & $\bar{\pi}_{g_i}$ & $\lvert z\rvert$\\
	\colrule	
		
   	Line  		& 2 & 2520 & 0.1499375 	& 0.1499594(157) & 1.39\\
	Fork   		& 2 & 2520 & 0.1499375 	& 0.1499275(131) & 0.77\\
	Trident   	& 6 & 840 & 0.0499792 	& 0.0499930(083) & 1.69\\
	Pitchfork & 24 & 210 & 0.0124948 	& 0.0124985(044) & 0.84\\
	Star  		& 720 & 7 & 0.0004165 	& 0.0004170(007) & 0.84\\
	Handle  	& 1 & 5040 & 0.2998751 	& 0.2998690(170) & 0.36\\
	HandleFork & 2 & 2520 & 0.1499375	& 0.1499251(148) & 0.84\\
	Pentane 	& 8 & 630 & 0.0374843 	& 0.0374749(070) & 1.35\\
	TriFork	& 12 & 420 & 0.0249896 	& 0.0249928(056) & 0.57\\
	Tri 	& 6 & 840 & 0.0499792 	& 0.0499719(115) & 0.63\\
	Cross & 4 & 1260 & 0.0749688 	& 0.0749710(123) & 0.18\\

	\end{tabular}
	\end{ruledtabular}
	\centering
	\vspace{5mm}	
	\caption{Sampled ($\bar{\pi}_i$) and exact ($\pi_i$) isomorphism class probability distribution of the $K_7$ ensemble. Over $70\%$ of the $z$ statistics are within one standard error and all are within two. $\chi^2 = 9.94$ for ten degrees of freedom. The probability of finding a $\chi^2$ as large as this is $P = 0.55$. This value of $\chi^2$, $P$ and the distribution of the $z$ statistics by the empirical normal distribution rule strongly indicate that $\bar{\pi}$ is sampled from $\pi$ and that the errors are well under control.}
	\label{table:k7_results}
\end{table*}

Not shown in this section, but included in the appendix are the graph isomorphism classes for the $K_4$, $K_5$ and $K_6$ graphs. Also in the appendix are the estimated graph probability distributions, their errors and exact values in Tabs.~\ref{table:k4_results}, \ref{table:k5_results} and \ref{table:k6_results}.

The sampled probability distributions are found by histogramming the graph configurations over $s$ samples, generated after performing a rewiring sweep. We perform $m$ independent Monte Carlo replica to estimate the distributions $\pi_{g_i}^r$. The mean over these replica is taken and we use the bootstrap method to estimate the statistical error on this mean using $B$ bootstrap resamples. The error in the three least significant digits is presented in parentheses. The $z$-score shown in Tab.~\ref{table:k7_results} and defined as,
\begin{align*}
\lvert z \rvert = \bigg| \frac{\pi_{g_i} - \bar{\pi}_{g_i}}{\sigma_{\bar{\pi}_{g_i}}} \bigg|,
\end{align*}
illustrates how far $\bar{\pi}_{g_i}$ is from the exact distribution $\pi_{g_i}$ in units of the standard error.

To estimate $\bar{\pi}_{g_i}$, the rewiring algorithm was run for $s = 10^7$ sweeps, where one sweep involves $\lvert V \rvert$ rewires. This was repeated for $m = 100$ simulations. The statistical standard error of the mean, which are shown in parentheses in Tab.~\ref{table:k7_results}, are estimated using $B = 10^6$ bootstrap resamples of the $m$ independent probability distribution estimates.

The goodness of fit of $\bar{\pi}_{g_i}$ to the exact distribution $\pi_{g_i}$ can be found using the $\chi^2$ test. The $\chi^2$ statistic in our case is,
\begin{align}\label{rewiring:chi2_pi}
\chi^2 = \frac{1}{N}\sum_{i=g_{1}}^{g_N}  \left(\frac{\pi_{i} - \bar{\pi}_{i}}{\sigma_{\bar{\pi}_i}}\right)^2.
\end{align}
Our null hypothesis is that the observed $\bar{\pi}_{g_i}$ is consistent with $\pi_{g_i}$ and that any deviation between the two is purely by chance. This assumption is tested using $P$-values. The $P$-value is the probability of observing a $\chi^2$ statistic as large as what we have calculated. For significance, we assume that if $P < 0.05$, then $\bar{\pi}_{g}$ and its standard error are very unlikely to be sampling from $\pi_{g}$.

As we can see in the caption of Tab.~\ref{table:k7_results}, the $\chi^2 = 9.94$ for ten degrees of freedom and $70\%$ of $z$-scores are within one standard error. The estimated probability distribution is in very good agreement with the exact values and we can be confident that our rewiring algorithm is sampling from the distribution of $K_7$ correctly.

This evidence suggests that our algorithm can accurately estimate the graph probability distribution once the ensemble is small enough to identify every graph isomorphism class. The errors in the graph probability distribution can also be estimated with good precision. These results give us confidence in our algorithm and naturally lead to estimating the mean of graph observables.

\subsection{Graph Diameter} \label{rewiring:graph_diameter}

Having demonstrated that we can accurately and precisely reproduce the graph distribution for small graph ensembles, the next step is to use our algorithm to estimate ensemble averages. The graph diameter was chosen as an interesting graph observable for a number of reasons; the diameter can be easily calculated on large graphs, furthermore, it depends on the entire graph structure and therefore changes slowly due to local updates and thus serves as a possible probe for $\tau_{\mathrm{exp}}$, the largest autocorrelation time. Finally, an asymptotic value related to the graph diameter expectation value in the limit as graph size tends towards infinity exists to which we can compare our Monte Carlo estimate~\cite{Szekeres:67}.

In order to estimate the mean graph diameter with a small standard error, we first show that the integrated autocorrelation time ($\tau_{\mathrm{int}}$) is under control. Then we demonstrate that we can accurately estimate the graph diameter which agrees with the expected value of the graph diameter for small graph ensembles. Finally, we show our estimate of the fit of the mean graph diameter from our Monte Carlo experiments agrees with the bound on the height of trees given in Eq.~\ref{eq: tree_height_bound}.

\subsubsection{Integrated Autocorrelation Time} \label{rewiring:autocorrelation}
The maximum graph diameter grows linearly with the graph size and changes slowly during graph rewires due to the local updating. It also gives a good description of the graph configuration. Estimating $\tau_{\mathrm{int}}$ of the diameter allows us to describe the efficiency of our algorithm in selecting structurally different graph configurations. To estimate $\tau_{\mathrm{int}}$, we used the method described in Ref.~\cite{Wolff:04}. The rewiring Monte Carlo algorithm was run for each graph ensemble and a total of $10^6$ graph diameter values were calculated. To control for thermalisation errors, $10^5$ rewiring sweeps were performed before starting the measurement phase.

\begin{figure}[H]
\centering
\includegraphics[width=\columnwidth]{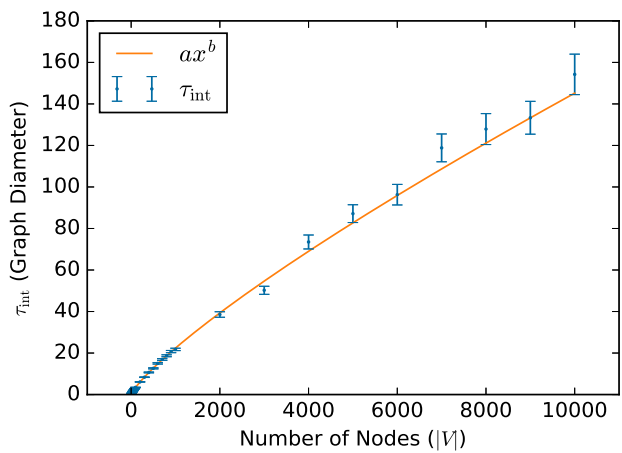}
\caption{Plot of the integrated autocorrelation time ($\tau_{\mathrm{int}}$) of the graph diameter versus number of nodes $\lvert V \rvert$, fitted to the monomial $\tau_{\mathrm{int}}(\lvert V \rvert) =  0.08233(76)\lvert V \rvert^{ 0.8116(21)}$ over three orders of magnitude. The fit uses all data points except the outlier at $\lvert V\rvert = 10$. The reduced $\chi^2$ for 24 degrees of freedom is 0.93.}
\label{fig:ch5_tau_int_scaling}
\end{figure}

Fig.~\ref{fig:ch5_tau_int_scaling} shows the estimated $\tau_{\mathrm{int}}$ values and their error bars over three orders of magnitude of graph size. The line of best fit,
\begin{align*}
f(x, a, b) = ax^b,
\end{align*}
was calculated using the Maximum Likelihood Estimation method and minimizing,
\begin{align} \label{eq:chi-square-se}
\chi^2 = \sum_i \left(\frac{f(x_i, a, b) - y_i}{\sigma_i} \right)^2,
\end{align}
using the least squares method.

The ansatz was not chosen based on some theoretical result, but rather from a \textit{by-eye} initial fit and should only be used as a guideline. The reduced $\chi^2$ indicates that our data gives an excellent fit to our model. The resulting monomial depends on the size of the graph and the exponent is below one. Therefore, as the system size increases through three orders of magnitude, $\tau_{\mathrm{int}}$ is still very manageable. While it is not reasonable to directly make a comparison of our $\tau_{\mathrm{int}} \approx \lvert V\rvert^{0.81}$ with the relaxation time calculated in Ref.~\cite{Kannan:99} of $O(\lvert E \rvert^6)$, it does compare well with the $7.5\lvert E \rvert$ and $15\lvert E \rvert$ rewires necessary to produce independent graph configurations from Ref.~\cite{Ray:15} for graphs of $10^3$ nodes.

Having shown that $\tau_{\mathrm{int}}$ of the graph diameter is under control, we can now estimate the error for the graph diameter precisely. Each graph diameter mean in this section and the next was estimated using $10^6$ samples measured after a rewiring sweep. The Markov chain was thermalized using $10^5$ sweeps. When the Markov chain graph diameter history is inspected, the initial transient from low probability graph diameter configurations towards the mean takes many times less sweeps than this thermalisation time. The statistical analysis of the diameter mean and variance were performed using the binning method described in Ref.~\cite{Berg:04}, using a bin size of 1000, which is much larger than $\tau_{\mathrm{int}}$. This results in uncorrelated means of each of the bins. The standard error of the sampling distribution of the sample mean was estimated using these 1000 binned means. These means were resampled using the $10^6$ Bootstrap resamples and the standard deviation of these resamples was calculated.

Tab.~\ref{table:diameter_means_vs_exact} shows that our estimate of the graph diameter mean and standard error agree well with the exact diameter expectation value. The $z$-score for the $K_4$ and $K_5$ ensemble suggest that the statistical errors may be slightly underestimated, which is reflected in the reduced chi-square which is just larger than we would like. However, the results of the $K_6$ and $K_7$ mean diameters are in excellent agreement, suggesting that the larger $z$-scores should not be of too much concern.
\begin{table}[H]
	\centering
	\begin{ruledtabular}		
	\begin{tabular}{c c c c}
		Graph Ensemble & $\bar{d}$ & $\langle d \rangle$ & $\lvert z\rvert$\\
		\colrule
   		$K_4$	& 2.74944(44) 			& 2.75 & 1.29\\
		$K_5$	& 3.43920(53)			& 3.44 & 1.51\\
		$K_6$	& 4.10671(68)			& $4.106\dot{4}8\dot{1}$	& 0.33	\\
		$K_7$	& 4.71172(79)			& 4.711370	& 0.45	\\
	\end{tabular}
	\end{ruledtabular}
	\centering
	\vspace{5mm}	
	\caption{Monte Carlo ensemble average estimates of graph diameter with standard error on the mean vs. exact graph diameter mean values. The reduced chi-square for three degrees of freedom is $\chi_{\mathrm{red}}^2 = 1.4$. The dotted numbers indicate a repeating decimal.}
	\label{table:diameter_means_vs_exact}
\end{table}

The asymptotic behaviour of the mean height of trees as the number of nodes tend to infinity is given by Eq.~\ref{eq:expectation_height}. Eq.~\ref{eq: tree_height_bound} also shows that the height of trees is bounded from above by the graph diameter and from below by half the diameter. We therefore expect the form of the scaling behavior of the graph diameter in trees to be very similar.

\begin{figure}[H]
\centering
\includegraphics[width=\columnwidth]{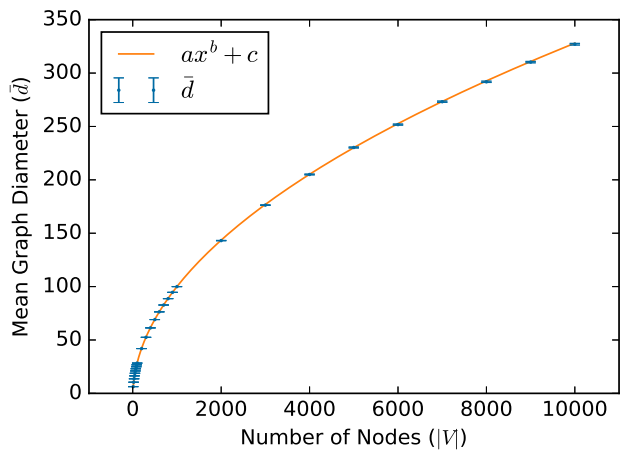}
\caption{Plot of the mean graph diameter $\bar{d}$ versus number of nodes $\lvert V \rvert$, fitted to the function $\bar{d}(\lvert V \rvert) =  3.2(1)\lvert V \rvert^{0.504(3)} - 4.0(9)$. The fit is performed on the data points where $\lvert V \rvert \geq 700.$ The reduced $\chi^2$ for 10 degrees of freedom is 1.02.}
\label{fig:ch5_graph_diameter_scaling}
\end{figure}

Fig.~\ref{fig:ch5_graph_diameter_scaling} shows our estimate of the graph diameter depending on $|V|$ as points with error bars and a continuous fitted model function based on the expected graph height. As expected, the close agreement between our numerical results and the model break down for small graph ensembles. With this in mind, we performed our fit using the data points for $\lvert V \rvert \geq 700$, however the fit line is plotted down to $\lvert V \rvert = 10$ using the fit parameters. Even at small graph sizes, where the fit breaks down, there is good agreement between the data and model function. The model function that we used is
\begin{align*}
f(x, a, b, c) = ax^b + c, 
\end{align*}
which takes the same form as Eq.~\ref{eq:expectation_height}. Our chi-square statistic is calculated using the same form as Eq.~\ref{eq:chi-square-se}. The errors for each data point were estimated using the binning method and bootstrap resampling. The additive term in our fit function remains small.
The value of the fit parameter $b=0.504(3)$ is found to be very close to the exponent of the expected value of $0.5$. Our model results in a reduced chi-square of $1.02$, with the residuals between our data and the model randomly scattered within two standard deviations from the mean. To the best of our knowledge, our numerical results provide the most precise estimate of the mean of the graph diameter in spanning trees with $|V|$ nodes.

\section{Conclusions and Future Work \label{sec:conclusion}}
In this paper, we outline a novel rewiring algorithm which can be used to sample over the configuration space of spanning trees with $\lvert V \rvert$ nodes. The resulting Markov chain is proven to be ergodic, which makes it useful to calculate ensemble statistics. The graph probability distribution is estimated for small graphs and agrees very well with the exact distribution. 

The graph diameter, which describes the structure of graph configurations is then investigated. The integrated autocorrelation time of the graph diameter scales like $\tau_{\mathrm{int}} \approx \lvert V \lvert ^{0.81}$, which means that the estimates of the diameter mean and variance can be calculated even for large graphs. The form of the asymptotic expectation value of the height of trees and our numerical results agree and the graph diameter scales like $\bar{d} \approx \lvert V \rvert^{0.5}$. The goodness of fit, given by the reduced chi-square suggests that our choice of model is appropriate and validates the form of the height of trees expectation value. 

In terms of future work stemming from this paper, we see this as early work and endeavor to study graph ensembles for arbitrary graph types, not only spanning trees. An important contribution to this area would be the ability to rewire graphs with a flexible number of nodes and edges. Furthermore, the methods outlined in this paper can be used to study dynamic models embedded on dynamic networks. In many situations, it may be of interest to calculate ensemble averages over the joint probability distribution of graph and spin configurations. Finally, our approach allows the future study of more complicated structural properties and graph observables of spanning trees.

\section*{Acknowledgements}
This work is supported by Science Foundation Ireland through the CTVR and CONNECT grants 10/CE/i1853 and 13/RC/2077, respectively. Calculations were performed on the Lonsdale cluster maintained by the Trinity Centre for High Performance Computing. This cluster was funded through grants from Science Foundation Ireland.

\appendix* \section*{Appendix A: Graph probability distributions}

In this appendix, the Monte Carlo results of our rewiring algorithm are presented for the $K_4$, $K_5$ and $K_6$ spanning tree ensembles. The non-isomorphic spanning trees and a table of the exact and estimated probability distribution of spanning trees is shown in each case.

\subsection{\texorpdfstring{$K_4$ Ensemble}{K4 Ensemble}}
\begin{figure}[H]
	\centering
	\subfloat[Line]{
	\begin{tikzpicture} \label{graph:n4_line}
		\tikzstyle{every node} = [draw, shape=circle, minimum size = 0.5cm]
		\node(v1) at (0,1) {};
		\node(v2) at (1,1) {};
		\node(v3) at (2,1) {};
		\node(v4) at (3,1) {};
		\draw (v1) -- (v2);
		\draw (v2) -- (v3);
		\draw (v3) -- (v4);
	\end{tikzpicture}}\quad
	\subfloat[Star]{
	\begin{tikzpicture} \label{graph:n4_star}
		\tikzstyle{every node} = [draw, shape=circle, minimum size = 0.5cm]
		\node(v1) at (0,0) {};
		\node(v2) at (1,0) {};
		\node(v3) at (1.5,0.866) {};
		\node(v4) at (1.5,-0.866) {};
		\draw (v2) -- (v1);
		\draw (v2) -- (v3);
		\draw (v2) -- (v4);
	\end{tikzpicture}}\quad
	\caption{These graphs are the non-isomorphic spanning trees of $K_{4}$. All spanning trees of $K_4$ are isomorphic to one of these graphs.}
	\label{graph:K_4_spanning_trees}
\end{figure}
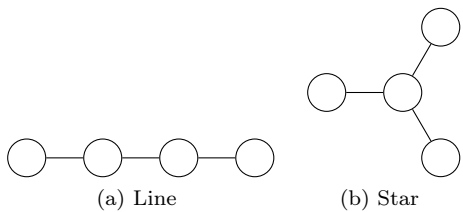

\begin{table}[H]
	\centering	
	\begin{ruledtabular}	
	\begin{tabular}{l c c c c c}
	Tree $(n^{n-2}=16)$ & $\lvert \mathrm{Aut}(g_i) \rvert$ & $l_{g_i}$ & $\pi_{g_{i}}$ & $\bar{\pi}_{g_i}$ & $\lvert z \rvert$\\
	\colrule
	Line (Fig.~\ref{graph:n4_line})  & 2 & 12 & 0.75 & .7500009(102) & 0.09\\
	Star (Fig.~\ref{graph:n4_star})  & 6 & 4 & 0.25 & .2499991(102) & 0.09\\
	\end{tabular}
	\end{ruledtabular}
	\centering
	\vspace{5mm}	
	\caption{The sampled ($\bar{\pi}_{g_i}$) and exact ($\pi_{g_i}$) isomorphism class probability distribution of the $K_4$ ensemble. The $z$-statistic shows that the sampled probability distribution is very close to the expected value with far less than one $\sigma$ in the difference. The $\chi^2 = 0.02$ for one degree of freedom, as described in Eq.~\ref{rewiring:chi2_pi}. The probability of finding a $\chi^2$ as large as this is $P = 0.11$. }	\label{table:k4_results}
\end{table}

\subsection{\texorpdfstring{$K_5$ Ensemble}{K5 Ensemble}}
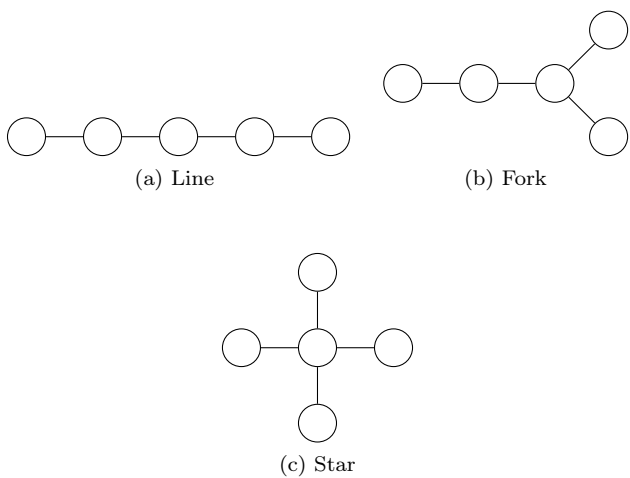
\begin{figure} [H]
	\centering
	\subfloat[Line]{
	\begin{tikzpicture} \label{graph:n5_line}
		\tikzstyle{every node} = [draw, shape=circle, minimum size = 0.5cm]
		\node(v1) at (0,0) {};
		\node(v2) at (1,0) {};
		\node(v3) at (2,0) {};
		\node(v4) at (3,0) {};
		\node(v5) at (4,0) {};
		\draw (v1) -- (v2);
		\draw (v2) -- (v3);
		\draw (v3) -- (v4);
		\draw (v4) -- (v5);
	\end{tikzpicture}}\quad
	\subfloat[Fork]{
		\begin{tikzpicture} \label{graph:n5_fork}
			\tikzstyle{every node} = [draw, shape=circle, minimum size = 0.5cm]
			\node(v1) at (0,0) {};
			\node(v2) at (1,0) {};
			\node(v3) at (2,0) {};
			\node(v4) at (2.707,0.707) {};
			\node(v5) at (2.707,-0.707) {};
			\draw (v1) -- (v2);
			\draw (v2) -- (v3);
			\draw (v3) -- (v4);
			\draw (v3) -- (v5);
		\end{tikzpicture}
	} \\ \bigskip
	\subfloat[Star]{
		\begin{tikzpicture} \label{graph:n5_star}
			\tikzstyle{every node} = [draw, shape=circle, minimum size = 0.5cm]
			\node(v1) at (0,0) {};
			\node(v3) at (1,0) {};
			\node(v4) at (0,-1) {};
			\node(v5) at (-1,0) {};
			\node(v2) at (0,1) {};
			\draw (v1) -- (v2);
			\draw (v1) -- (v3);
			\draw (v1) -- (v4);
			\draw (v1) -- (v5);
		\end{tikzpicture}
	}
	\caption{Non-isomorphic spanning trees of $K_{5}$.}\label{graph:K_5_spanning_trees}
\end{figure}

\begin{table} [H]
	\centering
	\begin{ruledtabular}		
	\begin{tabular}{c c c c c c}
		Tree $(n^{n-2}=125)$ & $\lvert \mathrm{Aut}(g_i) \rvert$ & $l_{g_i}$ & $\pi_{g_{i}}$ & $\bar{\pi}_{g_i}$ & $\lvert z \rvert$\\
		\colrule
   		Line (Fig.~\ref{graph:n5_line})  & 2 & 60 & 0.48 & .4800174(177) & 0.98\\
		Fork (Fig.~\ref{graph:n5_fork})  & 2 & 60 & 0.48 & .4799775(151) & 1.62\\
		Star (Fig.~\ref{graph:n5_star})  & 24 & 5 & 0.04 & .0400051(061) & 0.82\\
	\end{tabular}
	\end{ruledtabular}
	\centering
	\vspace{5mm}	
	\caption{Sampled ($\bar{\pi}_i$) and exact ($\pi$) isomorphism class probability distribution of the $K_5$ ensemble. Two thirds of the $z$ statistics are within one standard deviation, indicating that the $\bar{\pi}$ means are normally distributed and that the errors are reasonable. $\chi^2 = 3.87$ for two degrees of freedom. The probability of finding a $\chi^2$ as large as this is $P = 0.86$. The $\chi^2$ is a bit large, but there is still a good likelihood that $\bar{\pi}$ is sampled from $\pi$.}
	\label{table:k5_results}
\end{table}

\subsection{\texorpdfstring{$K_6$ Ensemble}{K6 Ensemble}}
\begin{table*}
	\centering		
	\begin{ruledtabular}
	\begin{tabular}{l c c c c c}
		Tree $(n^{n-2}=1296)$ & $\lvert \mathrm{Aut}(g_i) \rvert$ & $l_{g_i}$ & $\pi_{g_{i}}$ & $\bar{\pi}_{g_i}$ & $\lvert z\rvert$\\
		\colrule
   		Line (Fig.~\ref{graph:n6_line})  	& 2 & 360 & $0.2\dot{7}$ & .2777699(204) & 0.39\\
		Fork (Fig.~\ref{graph:n6_fork})  	& 2 & 360 & $0.2\dot{7}$ & .2777847(137) & 0.5\\
		Trident (Fig.~\ref{graph:n6_trident})& 6 & 120 & 0.0925926 & .0925895(111) & 0.27\\
		Star (Fig.~\ref{graph:n6_star})  	& 120 & 6 & 0.0046296 & .0046303(023) & 0.29\\
		Butane (Fig.~\ref{graph:n6_Butane})  & 8 & 90 & $0.069\dot{4}$ & .0694618(080) & 2.17\\
		Handle (Fig.~\ref{graph:n6_Handle})  & 2 & 360 & $0.2\dot{7}$ & .2777638(200) & 0.7\\
	\end{tabular}
	\end{ruledtabular}
	\centering
	\vspace{5mm}	
	\caption{Sampled ($\bar{\pi}_i$) and exact ($\pi$) isomorphism class probability distribution of the $K_6$ ensemble. The majority of the $z$ statistics are within one standard deviation, however $\bar{\pi}$ of the Butane class is larger than we would expect to see randomly among six categories. $\chi^2 = 5.77$ for five degrees of freedom. The probability of finding a $\chi^2$ as large as this is $P = 0.66$. The $\chi^2$ is slightly big, but still reasonable. Considering that the $\chi^2$ statistic is extremely sensitive to outliers, this shows a good likelihood that $\bar{\pi}$ is sampled from $\pi$.}
	\label{table:k6_results}
\end{table*}

\begin{figure} [H]
	\centering
	\subfloat[Line]{
	\begin{tikzpicture} \label{graph:n6_line}
		\tikzstyle{every node} = [draw, shape=circle, minimum size = 0.5cm]
		\node(v1) at (0,0) {};
		\node(v2) at (1,0) {};
		\node(v3) at (2,0) {};
		\node(v4) at (3,0) {};
		\node(v5) at (4,0) {};
		\node(v6) at (5,0) {};

		\draw (v1) -- (v2);
		\draw (v2) -- (v3);
		\draw (v3) -- (v4);
		\draw (v4) -- (v5);
		\draw (v5) -- (v6);
	\end{tikzpicture}}\\ \bigskip
	\subfloat[Fork]{
		\begin{tikzpicture} \label{graph:n6_fork}
			\tikzstyle{every node} = [draw, shape=circle, minimum size = 0.5cm]
			\node(v1) at (0,0) {};
			\node(v2) at (1,0) {};
			\node(v3) at (2,0) {};
			\node(v4) at (3,0) {};
			\node(v5) at (3.707,0.707) {};
			\node(v6) at (3.707,-0.707) {};
			\draw (v1) -- (v2);
			\draw (v2) -- (v3);
			\draw (v3) -- (v4);
			\draw (v4) -- (v5);
			\draw (v4) -- (v6);
		\end{tikzpicture}
	} \\
	\bigskip
	\subfloat[Trident]{
		\begin{tikzpicture} \label{graph:n6_trident}
			\tikzstyle{every node} = [draw, shape=circle, minimum size = 0.5cm]
			\node(v1) at (0,0) {};
			\node(v2) at (1,0) {};
			\node(v3) at (2,0) {};
			\node(v4) at (2.707,0.707) {};
			\node(v5) at (3,0) {};
			\node(v6) at (2.707,-0.707) {};
			\draw (v1) -- (v2);
			\draw (v2) -- (v3);
			\draw (v3) -- (v4);
			\draw (v3) -- (v5);
			\draw (v3) -- (v6);
		\end{tikzpicture}
	} \quad
	\subfloat[Star]{
		\begin{tikzpicture} \label{graph:n6_star}
			\tikzstyle{every node} = [draw, shape=circle, minimum size = 0.5cm]
			\node(v1) at (0:0) {};
			\node(v2) at (90:1) {};
			\node(v3) at (162:1) {};
			\node(v4) at (232:1) {};
			\node(v5) at (304:1) {};
			\node(v6) at (16:1) {};
			\draw (v1) -- (v2);
			\draw (v1) -- (v3);
			\draw (v1) -- (v4);
			\draw (v1) -- (v5);
			\draw (v1) -- (v6);
		\end{tikzpicture}
	}\\
	\bigskip
		\subfloat[Butane]{
		\begin{tikzpicture} \label{graph:n6_Butane}
			\tikzstyle{every node} = [draw, shape=circle, minimum size = 0.5cm]
			\node(v1) at (0,0.707) {};
			\node(v2) at (0,-0.707) {};
			\node(v3) at (0.707,0) {};
			\node(v4) at (1.707,0) {};
			\node(v5) at (2.414,0.707) {};
			\node(v6) at (2.414,-0.707) {};
			\draw (v1) -- (v3);
			\draw (v2) -- (v3);
			\draw (v3) -- (v4);
			\draw (v4) -- (v5);
			\draw (v4) -- (v6);
		\end{tikzpicture}
	} \quad
	\subfloat[Handle]{
	\begin{tikzpicture} \label{graph:n6_Handle}
		\tikzstyle{every node} = [draw, shape=circle, minimum size = 0.5cm]
		\node(v1) at (0,0) {};
		\node(v2) at (1,0) {};
		\node(v3) at (2,0) {};
		\node(v4) at (3,0) {};
		\node(v5) at (4,0) {};
		\node(v6) at (2,1) {};

		\draw (v1) -- (v2);
		\draw (v2) -- (v3);
		\draw (v3) -- (v4);
		\draw (v4) -- (v5);
		\draw (v3) -- (v6);
	\end{tikzpicture}}
	\caption{Non-isomorphic spanning trees of $K_{6}$.}\label{graph:K_6_spanning_trees}
\end{figure}
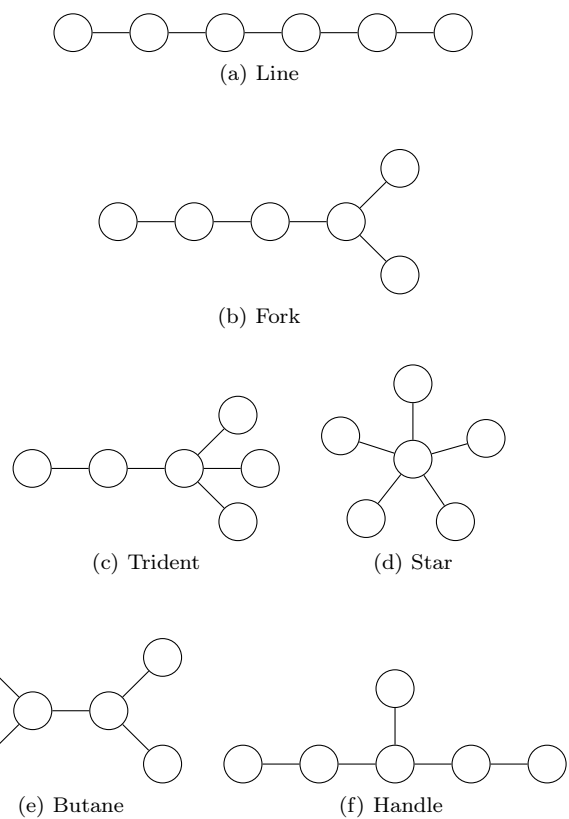

\end{document}